\documentclass[12pt,english]{article}
\usepackage[latin9]{inputenc}
\usepackage[letterpaper]{geometry}
\usepackage{xcolor}
\usepackage{pdfcolmk}
\usepackage{array}
\usepackage{calc}
\usepackage{multirow}
\usepackage{amsthm}
\usepackage{amsmath}
\usepackage{graphicx}
\usepackage[all]{xy}
\PassOptionsToPackage{normalem}{ulem}
\usepackage{ulem}

\makeatletter

\providecommand{\tabularnewline}{\\}
\providecolor{lyxadded}{rgb}{0,0,1}
\providecolor{lyxdeleted}{rgb}{1,0,0}

\newcommand{\lyxaddress}[1]{
\par {\raggedright #1
\vspace{1.4em}
\noindent\par}
}

\usepackage[all]{xy}

\usepackage{cite}

\usepackage{placeins}

\usepackage{lineno}

\newcommand{\sodd}{\mathsf{s}_{1}}

\def\frontmatter@abstractheading{}

\date{}

\usepackage{pdfpages}

\makeatother

\usepackage{babel}
\begin{document}

\title{Is there contextuality in behavioral and social systems?}

\author{Ehtibar N. Dzhafarov\textsuperscript{1}, Ru Zhang\textsuperscript{1}
and Janne Kujala\textsuperscript{2}}

\maketitle

\lyxaddress{\begin{center}
\textsuperscript{1}Purdue University, ehtibar@purdue.edu \\\textsuperscript{2}University
of Jyv\"askyl\"a, jvk@iki.fi
\par\end{center}}
\begin{abstract}
Most behavioral and social experiments aimed at revealing contextuality
are confined to cyclic systems with binary outcomes. In quantum physics,
this broad class of systems includes as special cases Klyachko-Can-Binicioglu-Shumovsky-type,
Einstein-Podolsky-Rosen-Bell-type, and Suppes-Zanotti-Leggett-Garg-type
systems. The theory of contextuality known as Contextuality-by-Default
allows one to define and measure contextuality in all such system,
even if there are context-dependent errors in measurements, or if
something in the contexts directly interacts with the measurements.
This makes the theory especially suitable for behavioral and social
systems, where direct interactions of ``everything with everything''
are ubiquitous. For cyclic systems with binary outcomes the theory
provides necessary and sufficient conditions for noncontextuality,
and these conditions are known to be breached in certain quantum systems.
We review several behavioral and social data sets (from polls of public
opinion to visual illusions to conjoint choices to word combinations
to psychophysical matching), and none of these data provides any evidence
for contextuality. Our working hypothesis is that this may be a broadly applicable
rule: behavioral and social systems are noncontextual, i.e., all ``contextual
effects'' in them result from the ubiquitous dependence of response
distributions on the elements of contexts other than the ones to which
the response is presumably or normatively directed. 

\end{abstract}
\markboth{}{Dzhafarov, Zhang, Kujala}

\section{\label{sec:Introduction}Introduction}

Although the word is widely used in linguistics, psychology, and philosophy,
the notion of contextuality as it is used in this paper comes from
quantum mechanics, where in turn it came from logic \cite{specker1960}.
The reason for the prominence of this notion in quantum theory is
that classical-mechanical systems are not contextual while some quantum-mechanical
systems are. Contextuality is sometimes even presented as one of the
``paradoxes'' of quantum mechanics. In psychology, as it turns out,
a certain variety of (non)contextuality has been prominent too, but
it is known under different name: selectiveness of influences, or
lack thereof\emph{ }(for details, see Refs. \cite{DK2012JMP,DK2012LNCS}). 

The term ``contextuality'' refers to properties of systems of random
variables each of which can be viewed (sometimes artificially) as
a measurement of some ``object'' in some \emph{context}. For instance,
an object $q$ may be a question, and the context may be defined by
what other question $q'$ it is asked in combination with. Then the
answer to this question is a random variable $R_{q}^{(q,q')}$
that can be interpreted as the measurement of $q$ in the context
$(q,q')$. If the same question $q$ is then asked in combination
with some other question $q''$, then the measurement is a different
random variable, $R_{q}^{(q,q'')}$. More generally, context
in which $q$ is measured is defined by the conditions $c$ under
which the measurement is made, yielding random variable $R_{q}^{c}$.
This notation (or one of numerous variants thereof) is called \emph{contextual
notation} for random variables: it codifies the idea that the identity
of a measurement is defined both by what is measured and by the conditions
under it is measured \cite{Khr2005,Khr2009,Svozil,Winter2014,DK2013PLOS,DK2014Scripta,DKL2015LNCS,Larsson2002}. 

Within each context the measurements are made ``together'', because
of which they have an empirically defined \emph{joint distribution}.
Thus, in context $(q,q')$ we have two jointly distributed
random variables $R_{q}^{(q,q')}$ and $R_{q'}^{(q,q')}$.
We call the set of all random variables jointly recorded in a given
context a \emph{bunch} (of random variables, or of measurements).
Two different bunches have no joint distribution, because there is
no empirically defined way of coupling the values of one bunch with
those of another. We say that they are \emph{stochastically unrelated}.
Thus, in 
\begin{equation}
R^{(q,q')}=(R_{q}^{(q,q')},R_{q'}^{(q,q')})\textnormal{ and }R^{(q,q'')}=(R_{q}^{(q,q'')},R_{q''}^{(q,q'')})\label{eq:2 bunches}
\end{equation}
any component of $R^{(q,q')}$ is stochastically unrelated
to any component of $R^{(q,q'')}$, including $R_{q}^{(q,q')}$
and $R_{q}^{(q,q'')}$.

This work is based on the theory of contextuality dubbed Contextuality-by-Default
(CbD) \cite{KDconjecture,DK2015,KD2015,DKL2015,DKL2015LNCS,KDL2015,DK2014Scripta,DK_PLOS_2014}
(for precursors of this theory, see Refs. \cite{Larsson2002,Svozil,Winter2014}).
On a very general level, its main idea is that 
\begin{quote}
\emph{a system of different, stochastically unrelated bunches of random
variables can be characterized by considering all possible ways in
which they can be coupled under well-chosen constraint}s\emph{ imposed,
for each object, on the relationship between the measurements of this
object in different contexts}. 
\end{quote}
To \emph{couple} different bunches simply means to impose a joint
distribution on them. In the example above, this means finding four
jointly distributed random variables $(A,B,X,Y)$ such
that, in reference to (\ref{eq:2 bunches}), 
\begin{equation}
(A,B)\sim R^{(q,q')}\textnormal{ and }(X,Y)\sim R^{(q,q'')},
\end{equation}
$\sim$ standing for ``is distributed as''. The quadruple $(A,B,X,Y)$
is then called a \emph{coupling} for the bunches $R^{(q,q')}$
and $R^{(q,q'')}$. The ``well-chosen constraints'' is
a key notion in the formulation above. In our example, these constraints
should apply to $A$ and $X$, the coupling counterparts of $R_{q}^{(q,q')}$
and $R_{q}^{(q,q'')}$ measuring (answering) the same question
$q$ in two different contexts. 

Intuitively, ``noncontextuality'' means ``independence of context'',
and because of this it is tempting to say that the system of two bunches
in (\ref{eq:2 bunches}) is noncontextual if we can consider $R_{q}^{(q,q')}$
and $R_{q}^{(q,q'')}$ as ``one and the same'' random
variable, $R_{q}$. This may appear simple, but in fact it is logically
impossible: since $R_{q}^{(q,q')}$ and $R_{q}^{(q,q'')}$
are stochastically unrelated, they cannot be ``the same''. A random
variable cannot be stochastically unrelated to itself. The precise
meaning here comes from considering couplings $(A,B,X,Y)$
for the two bunches. Clearly, in every such a coupling $A\sim R_{q}^{(q,q')}$
and $X\sim R_{q}^{(q,q'')}$. We can say that the measurement
of $q$ in the system is context-independent if among all possible
couplings $(A,B,X,Y)$ there is at least one in which $\Pr[A\not=X]=0$.
In this particular example, due to its simplicity (only three random
variables involved in two contexts) it can be shown that such a coupling
does exist, provided $R_{q}^{(q,q')}\sim R_{q}^{(q,q'')}$.
In a more complex system, such a coupling may not exists even if the
system is \emph{consistently connected}: which means that in this
system the measurements of one and the same ``object'' always have
the same distribution.

The traditional approaches to contextuality were confined to consistent
connectedness, but this condition is too restrictive in quantum physics
\cite{bacciagaluppi,DKL2015,KDL2015} and virtually inapplicable in
social and behavioral sciences: almost always, a response to question
(or stimulus) $q$ will depend on the context in which it is asked,
which may translate into $R_{q}^{(q,q')}$ and $R_{q}^{(q,q'')}$
having different distributions. There is nothing wrong in calling
any such a case contextual, and this is done by many (see Sections
\ref{sec:Question-order-effect} and \ref{sec:Word-combinations:-Results}
below). It is, however, more informative to separate inconsistent
connectedness from contextuality, and this is what is done in the
CbD theory. We use the term \emph{inconsistently connected} for the
systems that are not necessarily consistently connected (but may be
so, as a special or limit case). 

The logic of the CbD approach is as follows. We first consider separately
the random variables measuring the same object in different contexts,
in our example $R_{q}^{(q,q')}$ and $R_{q}^{(q,q'')}$.
We call this set of random variables the \emph{connection} (for the
measured object, in our case $q$). Among all possible couplings $(A',X')$
for the connection $\{ R_{q}^{(q,q')},R_{q}^{(q,q'')}\} $,
i.e., among all jointly distributed $(A',X')$ such that
$A'\sim R_{q}^{(q,q')}$ and $X'\sim R_{q}^{(q,q'')}$,
we find the minimal value $m'$ of $\Pr[A'\not=X']$. Then
we look at the entire system of the bunches, in our case (\ref{eq:2 bunches}),
and among all possible couplings $(A,B,X,Y)$ for this
system we find the minimal value $m$ for $\Pr[A\not=X]$.
It should be clear that $m'$ cannot exceed $m$, because in every
coupling $(A,B,X,Y)$ for (\ref{eq:2 bunches}) the part
$(A,X)$ forms a coupling for the connection $\{ R_{q}^{(q,q')},R_{q}^{(q,q'')}\} $.
But they can be equal, $m=m'$, and then we say that the system is
noncontextual. If $m>m'$, the system is contextual. Again, due to
its simplicity, the system consisting of the two bunches (\ref{eq:2 bunches})
cannot be contextual, but this may very well be the case in more complex
systems. 

As an example of the latter, consider a system with two bunches
\begin{equation}
R^{(q,q')}=(R_{q}^{(q,q')},R_{q'}^{(q,q')})\textnormal{ and }R^{(q',q)}=(R_{q}^{(q',q)},R_{q'}^{(q',q)})\label{eq:2 bunches order}
\end{equation}
in which there are only two ``objects'' $q,q'$, and the two contexts
differ in the order in which these objects are measured. We have two
connections here,
\begin{equation}
\{ R_{q}^{(q,q')},R_{q}^{(q',q)}\} \textnormal{ and }\{ R_{q'}^{(q,q')},R_{q'}^{(q',q)}\} .
\end{equation}
Let us assume the measurements are binary, with values $+1$ and $-1$
(e.g., corresponding to answers Yes and No), and let us further assume
that all four random variables are ``fair coins'', with equal probabilities
of +1 and -1. Then the distribution of the bunches $R^{(q,q')}$
and $R^{(q',q)}$ in (\ref{eq:2 bunches order}) are uniquely
defined by the product expected values $\langle R_{q}^{(q,q')}R_{q'}^{(q,q')}\rangle $
and $\langle R_{q}^{(q',q)}R_{q'}^{(q',q)}\rangle $. 

It easy to see that, across all possible couplings $(A',X')$
for $\{ R_{q}^{(q,q')},R_{q}^{(q',q)}\} $,
the minimum value $m'_{1}$ of $\Pr[A'\not=X']$ is 0,
and the same is true for the minimum value $m'_{2}$ of $\Pr[B'\not=Y']$
across all possible couplings $(B',Y')$ for $\{ R_{q'}^{(q,q')},R_{q'}^{(q',q)}\} $.
However, it follows from the general theory that across
all possible couplings $(A,B,X,Y)$ for the entire system
(\ref{eq:2 bunches order}) the values $m_{1}$ of $\Pr[A\not=X]$
and $m_{2}$ of $\Pr[B\not=Y]$ cannot be both zero unless
$\langle R_{q}^{(q,q')}R_{q'}^{(q,q')}\rangle =\langle R_{q}^{(q',q)}R_{q'}^{(q',q)}\rangle $.
The latter need not be the case: it may, e.g., very well be that $\langle R_{q}^{(q,q')}R_{q'}^{(q,q')}\rangle =1$
(perfect correlation) and $\langle R_{q}^{(q',q)}R_{q'}^{(q',q)}\rangle =-1$
(perfect anti-correlation). In this case $m_{1}+m_{2}\geq1$, whence
either $m_{1}>m'_{1}=0$ or $m_{2}>m'_{2}=0$, indicating that the
system is contextual. 

As we show in this paper, the general rule for a broad spectrum of
behavioral and social systems of measurements seems to be that \emph{they
are all noncontextual in the sense of CbD}.

\section{\label{sec:Cyclic-systems-of}Cyclic systems of arbitrary rank}

In this section and throughout the rest of the paper we assume that
all our measurements are binary random variables, with values $\pm1$.

We apply the logic of the CbD theory to systems in which all objects
are measured in pairs so that each object belongs to precisely two
pairs. We call such systems \emph{cyclic}, because we can enumerate
the objects in such a system $q_{1},\ldots,q_{n}$ and arrange them
in a cycle 
\begin{equation}
\xymatrix@C=1cm{q_{1}\ar[r] & q_{2}\ar[r] & \cdots\ar[r] & q_{n-1}\ar[r] & q_{n}\ar@/^{1pc}/[llll],}
\end{equation}
in which any two successive objects form a context. The number $n$
is referred to as the \emph{rank} of the system. Our last example
in the previous section is a cyclic system of rank 2, the smallest
possible. 

In accordance with our notation, each object $q_{i}$ in a cyclic
system is measured by two random variables: $R_{q_{i}}^{(q_{i},q_{i\oplus1})}$
and $R_{q_{i}}^{(q_{i\ominus1},q_{i})}$, where the operations
$\oplus$ and $\ominus$ are cyclic addition and subtraction (so that
$n\oplus1=1$ and $1\ominus1=n$). Since there are no other random
variables involved, we can simplify notation: we will denote $R_{q_{i}}^{(q_{i},q_{i\oplus1})}$,
measuring the first object in the context, by $V_{i}$, and $R_{q_{i}}^{(q_{i\ominus1},q_{i})}$,
measuring the second object in the context, by $W_{i}$. As a result
each bunch in a cyclic system has the form $(V_{i},W_{i\oplus1})$;
e.g., the bunch of measurements for $(q_{1},q_{2})$ is
$(V_{1},W_{2})$, for $(q_{n},q_{1})$ the bunch
is $(V_{n},W_{1})$, etc.

Now we can represent a cyclic system of measurements in the form of
a $V-W$ cycle:
\begin{equation}
\xymatrix@C=1cm{V_{1}\ar@{-}[r] & W_{2}\ar@{.}[r] & V_{2}\ar@{-}[r] & W_{3}\ar@{.}[r] & \cdots\ar@{.}[r] & V_{n}\ar@{-}[r] & W_{1}\ar@/^{1pc}/@{.}[llllll],}
\end{equation}
where solid lines indicate bunches (joint measurements) and point
lines indicate connections (measurements of the object in different
contexts). 

It is proved in Refs. \cite{DKL2015,KDL2015,KDconjecture} that such
a system is noncontextual if and only if its bunches satisfy the following
inequality:
\begin{equation}
\Delta C=\sodd(\langle V_{1}W_{2}\rangle ,\ldots,\langle V_{n-1}W_{n}\rangle ,\langle V_{n}W_{1}\rangle )-(n-2)-\sum_{i=1}^{n}|\langle V_{i}\rangle -\langle W_{i}\rangle |\leq0,\label{eq:criterion general}
\end{equation}
where $\langle \cdot\rangle $ denotes expected value,
and the $\sodd$-part is the maximum of all linear combinations $\pm\langle V_{1}W_{2}\rangle \pm\ldots\pm\langle V_{n-1}W_{n}\rangle \pm\langle V_{n}W_{1}\rangle $
with the proviso that the number of minuses is odd. Note that the
criterion is written entirely in terms of the expectations of $V_{i}$,
$W_{i}$ and of the products $V_{i},W_{i\oplus1}$ ($i=1,\ldots,n$).
This means that the information about a cyclic system we need can
be presented in the form of the diagram 

\begin{equation}
\xymatrix@C=1cm{{\scriptstyle \langle V_{1}\rangle }\ar@{-}[r]^{\langle V_{1}W_{2}\rangle } & {\scriptstyle \langle W_{2}\rangle }\ar@{.}[r] & {\scriptstyle \langle V_{2}\rangle }\ar@{-}[r]^{\langle V_{2}W_{3}\rangle } & \cdots\ar@{.}[r] & {\scriptstyle \langle V_{n}\rangle }\ar@{-}[r]^{\langle V_{n}W_{1}\rangle } & {\scriptstyle \langle W_{1}\rangle }\ar@/^{1pc}/@{.}[lllll].}
\label{eq:diagram}
\end{equation}
We will use such diagrams to discuss experimental data in the subsequent
sections.

This criterion of noncontextuality is generally breached by quantum-mechanical
systems. Thus, for consistently connected systems, for $n=3$, the
inequality reduces to Suppes-Zanotti-Leggett-Garg inequality \cite{SuppesZanotti1981,leggett_quantum_1985},
for $n=4$ it acquires the form of the Clauser-Horn-Shimony-Holt inequalities
for the Einstein-Podolsky-Rosen-Bell paradigm \cite{Bell1964,9CHSH,15Fine},
and for $n=5$ (with an additional constraint) it becomes what is
known as Klyachko-Can-Binicioglu-Shumovsky inequality \cite{Klyachko}.
All of them are predicted by quantum theory and supported by experiments
to be violated by some quantum-mechanical systems. For $n=3$, using
the criterion (\ref{eq:criterion general}), violations are also predicted
for inconsistently connected systems \cite{bacciagaluppi}; and for
$n=5$ violations of (\ref{eq:criterion general}) were demonstrated
experimentally \cite{Lapkiewicz2011} (as analyzed in Ref. \cite{KDL2015}).

By contrast, we find no violations of (\ref{eq:criterion general})
in all known to us behavioral and social experiments aimed at revealing
contextuality: $\Delta C$ never exceeds zero. In the subsequent sections
we demonstrate this ``failure to fail'' the noncontextuality criterion
on several experimental studies, for cyclic systems of rank 2, 3,
and 4.

\section{\label{sec:Question-order-effect}Question order effect (cyclic systems
of rank 2)}

Wang, Solloway, Shiffrin, and Busemeyer \cite{Wang} considered 73
polls in which two questions, $A$ and $B$ (playing the role of ``objects''
$q_{1},q_{2}$ being measured), were asked in two possible orders,
$A\rightarrow B$ and $B\rightarrow A$ (forming two contexts). The
possible answers to each question, random variables 
\begin{equation}
V_{1}=R_{A}^{A\rightarrow B},W_{2}=R_{B}^{A\rightarrow B},V_{2}=R_{B}^{B\rightarrow A},W_{1}=R_{A}^{B\rightarrow A},
\end{equation}
were binary: $+1$ (Yes) or $-1$ (No). For instance, in the Gallup
poll results used in Ref. \cite{Moore}, one pair of questions was
(paraphrasing)
\begin{quote}
$A$: Do you think many white people dislike black people?

$B$: Do you think many black people dislike white people?
\end{quote}
with the resulting estimates of joint and marginal probabilities

\begin{center}
\begin{tabular}{c|c|c}
\cline{2-2} 
$A\rightarrow B$ & Yes to $B$ & \tabularnewline
\hline 
\multicolumn{1}{|c|}{Yes to $A$} & $.3987$ & \multicolumn{1}{c|}{.4161}\tabularnewline
\hline 
 & .5599 & ${\scriptstyle N\doteq500}$\tabularnewline
\cline{2-2} 
\end{tabular}$\quad$$\quad$%
\begin{tabular}{c|c|c}
\cline{2-2} 
 & Yes to $B$ & $B\rightarrow A$\tabularnewline
\hline 
\multicolumn{1}{|c|}{.5391} & $.4012$ & \multicolumn{1}{c|}{Yes to $A$}\tabularnewline
\hline 
${\scriptstyle N\doteq500}$ & .4609 & \tabularnewline
\cline{2-2} 
\end{tabular}
\par\end{center}

\noindent We translate ``Yes to $A$'' into $V_{1}=1$ in $A\rightarrow B$
and into $W_{1}=1$ in $B\rightarrow A$; correspondingly, ``Yes
to $B$'' translates into $W_{2}=1$ in $A\rightarrow B$ and into
$V_{2}=1$ in $B\rightarrow A$. Using the notation (\ref{eq:diagram}),
we deal here with the system
\[
\vcenter{\xymatrix@C=1cm{{\scriptstyle {\scriptstyle \langle V_{1}\rangle }}\ar@{-}[r]_{\langle V_{1}W_{2}\rangle } & {\scriptstyle {\scriptstyle \langle W_{2}\rangle }}\ar@{.}[d]\\
{\scriptstyle \langle W_{1}\rangle }\ar@{.}[u] & {\scriptstyle \langle V_{2}\rangle }\ar@{-}[l]_{\langle V_{2}W_{1}\rangle }
}
}=\vcenter{\xymatrix@C=1cm{{\scriptstyle -.1678}\ar@{-}[r]_{.6428} & {\scriptstyle {\scriptstyle .1198}}\ar@{.}[d]\\
{\scriptstyle .0782}\ar@{.}[u] & {\scriptstyle -.0782}\ar@{-}[l]_{.6048}
}
}
\]
To make sure the calculations are clear, for any $\pm1$ random variables
$X,Y$, 
\[
\begin{array}{c}
\langle X\rangle =2\Pr[X=1]-1,\\
\langle XY\rangle =\Pr[X=Y]-\Pr[X\not=Y]=4\Pr[X=1,Y=1]-2\Pr[X=1]-2\Pr[Y=1]+1.
\end{array}
\]
The noncontextuality criterion (\ref{eq:criterion general}) for cyclic
systems of rank 2 specializes to the form
\begin{equation}
\Delta C=|\langle V_{1}W_{2}\rangle -\langle V_{2}W_{1}\rangle |-(|\langle V_{1}\rangle -\langle W_{1}\rangle |+|\langle V_{2}\rangle -\langle W_{2}\rangle |)\leq0.\label{eq:criterion n=00003D2}
\end{equation}
For the values in the diagram above, $\Delta C=-0.406$, so there
is no evidence the system is contextual. 

Ref. \cite{Wang} contains analysis of 73 such pairs of questions,
including 66 taken from PEW polls (with $N$ ranging from 125 to 927),
four taken from Gallup polls reported by Moore \cite{Moore} (with
$N$ about 500), and three pairs of questions with $N$ ranging from
106 to 305. (The data were kindly provided to us by the authors of
Ref. \cite{Wang}; our computations based of these data are shown
in supplementary file S1.)

The analysis is simplified if we accept the empirical regularity discovered
by Wang and Busemeyer \cite{Wang-Busemeyer} and convincingly corroborated
in Ref. \cite{Wang}: using our notation, the discovery is that for
vast majority of question pairs,
\begin{equation}
\langle V_{1}W_{2}\rangle =\langle V_{2}W_{1}\rangle ,\label{eq:QQ}
\end{equation}
while
\begin{equation}
|\langle V_{1}\rangle -\langle W_{1}\rangle |+|\langle V_{2}\rangle -\langle W_{2}\rangle |\not=0.\label{eq:question order}
\end{equation}
The last inequality is what traditionally called the question order
effect \cite{Moore}, and (\ref{eq:QQ}) is dubbed by Wang and Busemeyer
the \emph{quantum question} (QQ) equality. Wang and Busemeyer \cite{Wang-Busemeyer}
theoretically justify the QQ equality by positing that the process
of answering two successive questions can be modeled by orthogonally projecting a state vector $\psi$ twice in a succession in a Hilbert space.
Denoting the projectors corresponding to response Yes to the questions
$A$ and $B$ by $P$ and $Q$, respectively, we have $P^{2}=P$,
$Q^{2}=Q$. The orthogonal projectors corresponding to response No
to the same two questions are then $I-P$ and $I-Q$, with $I$ denoting
the identity operator. We have, for the question order $A\rightarrow B$,
\[
\frac{1+\langle V_{1}W_{2}\rangle }{2}=\Vert QP\psi\Vert ^{2}+\Vert (I-Q)(I-P)\psi\Vert ^{2}=\langle (PQP+(I-P)(I-Q)(I-P))\psi\,|\,\psi\rangle ,
\]
and it is readily shown that 
\[
PQP+(I-P)(I-Q)(I-P)=I-(P+Q)+(PQ+QP).
\]
As $P$ and $Q$ enter in this expression symmetrically, the expression
is precisely the same for
\[
\frac{1+\langle V_{2}W_{1}\rangle }{2}=\Vert PQ\psi\Vert ^{2}+\Vert (I-P)(I-Q)\psi\Vert ^{2}.
\]
The empirical QQ effect now follows from the assumption that the operators
$P,Q$ do not vary across respondents (being determined by the questions
alone), whereas the mixture of the initial states $\psi$ has the
same distribution in any two large groups of respondents. At the same
time, the question order effect follows from the fact that $\Vert QP\psi\Vert ^{2}$
is not the generally the same as $\Vert PQ\psi\Vert ^{2}$. 

The QQ equality trivially implies (\ref{eq:criterion n=00003D2}),
i.e., lack of contextuality. Therefore, to the extent the QQ equality
can be viewed as an empirical law (and Ref. \cite{Wang} demonstrates
this convincingly for 72 out of 73 question pairs), the criterion
of noncontextuality should be satisfied for any $\langle V_{1}\rangle ,\langle W_{1}\rangle ,\langle V_{2}\rangle ,\langle W_{2}\rangle $.
We can confirm and complement the statistical analysis presented in
Ref. \cite{Wang} of the 72 questions by pointing out that the overall
chi-square test of the equality (\ref{eq:QQ}) over all of them yields
$p>0.35$, $df=72$. 

The singled out pair of questions that violates the QQ equality is
taken from the Gallup poll study reported in Ref. \cite{Moore}: paraphrasing,
\begin{quote}
$A$: Should Pete Rose be admitted to the baseball hall of fame?

$B$: Should shoeless Joe Jackson be admitted to the baseball hall
of fame?
\end{quote}
Refs. \cite{Wang-Busemeyer,Wang} provide an explanation for why the
double-projection model should not apply to this particular pair of
questions, but we need not be concerned with it. The diagram of the
results for this pair is

\[
\vcenter{\xymatrix@C=1cm{{\scriptstyle {\scriptstyle \langle V_{1}\rangle }}\ar@{-}[r]_{\langle V_{1}W_{2}\rangle } & {\scriptstyle {\scriptstyle \langle W_{2}\rangle }}\ar@{.}[d]\\
{\scriptstyle \langle W_{1}\rangle }\ar@{.}[u] & {\scriptstyle \langle V_{2}\rangle }\ar@{-}[l]_{\langle V_{2}W_{1}\rangle }
}
}=\vcenter{\xymatrix@C=1cm{{\scriptstyle .3241}\ar@{-}[r]_{.6190} & {\scriptstyle -.2886}\ar@{.}[d]\\
{\scriptstyle -.0346}\ar@{.}[u] & {\scriptstyle .0780}\ar@{-}[l]_{.3162}
}
},
\]
and it is readily seen to violate the equality $\langle V_{1}W_{2}\rangle =\langle V_{2}W_{1}\rangle $
($p<10^{-7}$, chi-square test with $df=1$). At the same time the
diagram yields $\Delta C=-0.422$, no evidence of contextuality. This
example serves as a good demonstration for the fact that while the
QQ equality is a sufficient condition for lack of contextuality, it
is by no means necessary.

Considering the question pairs one by one, all but six $\Delta C$
values out of 73 are negative. In five of these six cases, the QQ
equality $|\langle V_{1}W_{2}\rangle -\langle V_{2}W_{1}\rangle |=0$
cannot be rejected with $p$-values ranging from $0.06$ to $0.47$.
Therefore (\ref{eq:criterion n=00003D2}) cannot be rejected either.
In the remaining case, $p$-value for the QQ equality is $0.008$,
and $\Delta C=0.063$. While this case is suspicious, we do not think
it warrants a special investigation: using conventional significance
values, say, 0.01, for 73 similar cases we get the probability of
at least one rejection inflated to 0.52.

Note that in the literature cited, including Refs. \cite{Wang,Wang-Busemeyer},
the term ``contextual effect'' is used to designate the question
order effect (\ref{eq:question order}). This meaning of contextuality
corresponds to what we call here inconsistent connectedness (or violations
of marginal selectivity), and it should not be confused with the meaning
of contextuality as defined in Sections \ref{sec:Introduction} and
\ref{sec:Cyclic-systems-of} and indicated by the sign of $\Delta C$.

\section{Schr\"oder's staircase illusion (a cyclic system of rank 3)}

Asano, Hashimoto, Khrennikov, Ohya, and Tanaka \cite{Asano} studied
a cyclic system of rank 3, using as ``objects'' $q_{1},q_{2},q_{3}$
Shr\"oder's staircases tilted at three different angles, $\theta=40,45,50$
degrees, as shown in Figure 1. In fact, these three angles formed
the middle part of a set of 11 angles ranging from 0 to $90$ degrees
and presented either in the descending order (context $c_{1}$), or
in the ascending order (context $c_{2}$), or else in a random order
(context $c_{3}$). Each context involved a separate set of about
50 participants, and each participant in response to each of 11 angles
had to indicate whether she/he sees the surface A in front of B ($+1$)
or B in front of A ($-1$). From these 11 responses, in each context,
the authors selected two. In context $c_{1}$ the selected responses
where those to $\theta=40,45$ deg, so, formally, $c_{1}$ can be
identified with $(q_{1},q_{2})$; in contexts $c_{2}$
and $c_{3}$ the selected responses were those to $\theta=45,50$
deg and to $\theta=50,40$ deg, respectively, making $c_{2}=(q_{2},q_{3})$
and $c_{3}=(q_{3},q_{1})$. It is irrelevant to the logic
of the analysis that each context in fact contained all three tilts
$q_{1},q_{2},q_{3}$, as well as eight other tilts. (Ref. \cite{Asano}
includes a variety of other combinations of three objects and three
contexts extracted from the experiment in question. The data set for
the combination described here was kindly made available to us by
the authors of Ref. \cite{Asano}.)

\begin{figure}
\begin{centering}
\includegraphics[scale=0.25]{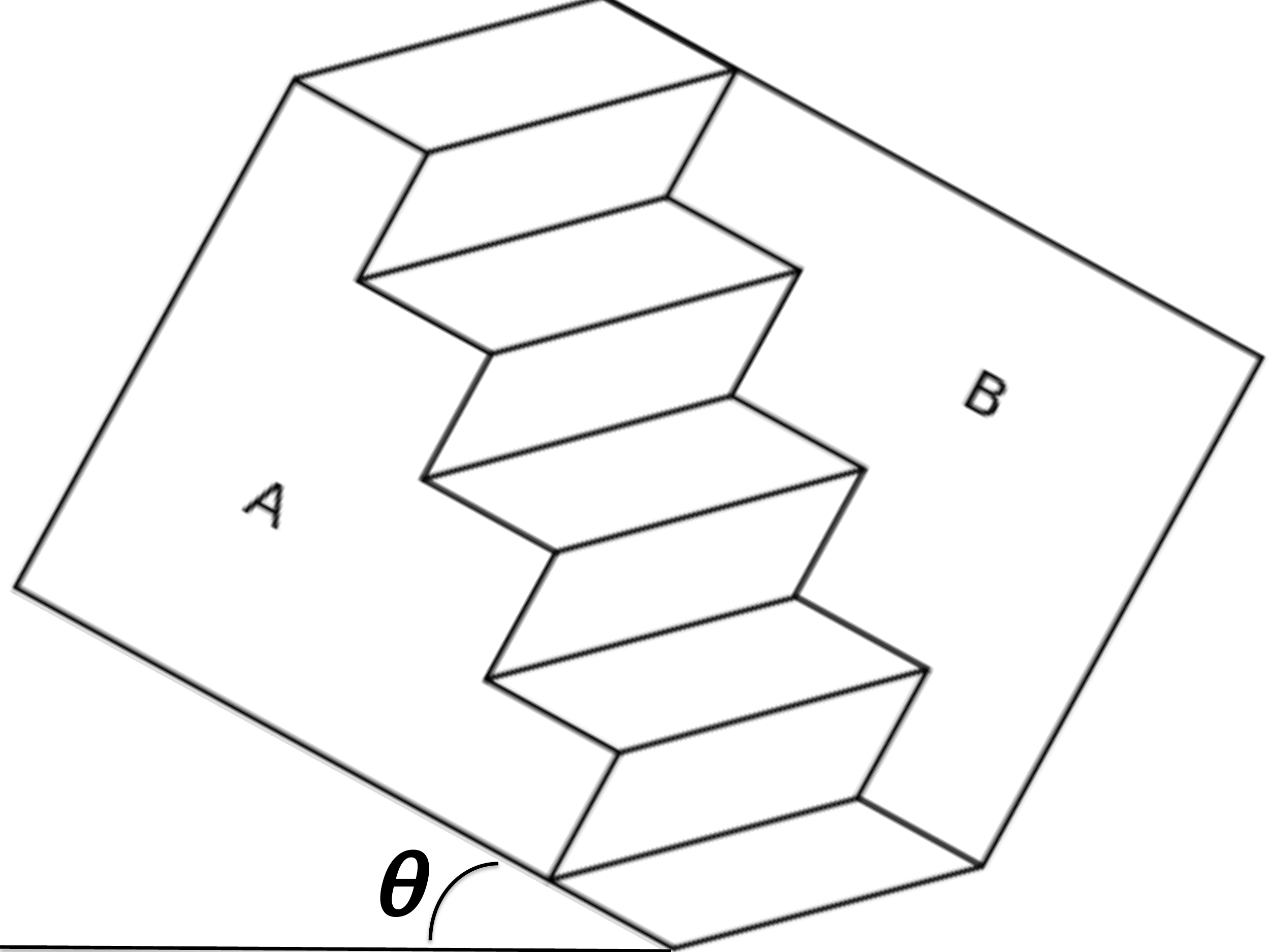}
\par\end{centering}

\protect\caption{Shr\"oder's staircases used in the experiments reported in Ref. \cite{Asano}}
\end{figure}

The results of the experiment are shown in the diagram of expected
values below: 

\[
\vcenter{\xymatrix@C=1cm{ & {\scriptstyle {\scriptstyle \langle V_{1}\rangle }}\ar@{-}[r]_{\langle V_{1}W_{2}\rangle } & {\scriptstyle \langle W_{2}\rangle }\ar@{.}[dr]\\
{\scriptstyle \langle W_{1}\rangle }\ar@{.}[ur] &  &  & {\scriptstyle \langle V_{2}\rangle }\ar@{-}[dl]_{\langle V_{2}W_{3}\rangle }\\
 & {\scriptstyle \langle V_{3}\rangle }\ar@{-}[ul]_{\langle V_{3}W_{1}\rangle } & {\scriptstyle \langle W_{3}\rangle }\ar@{.}[l]
}
}=\vcenter{\xymatrix@C=1cm{ & {\scriptstyle .708}\ar@{-}[r]_{.625} & {\scriptstyle .417}\ar@{.}[dr]\\
{\scriptstyle .382}\ar@{.}[ur] &  &  & {\scriptstyle -.333}\ar@{-}[dl]_{.625}\\
 & {\scriptstyle -.345}\ar@{-}[ul]_{.127} & {\scriptstyle -.625}\ar@{.}[l]
}
}
\]
The criterion of noncontextuality for a rank 3 cyclic system has the
form

\begin{equation}
\Delta C=\sodd(\langle V_{1}W_{2}\rangle ,\langle V_{2}W_{3}\rangle ,\langle V_{3}W_{1}\rangle )-1-\sum_{i-1}^{3}|\langle V_{i}\rangle -\langle W_{i}\rangle |\leq0\label{eq:criterion n=00003D3}
\end{equation}
where 
\[
\sodd(x,y,z)=\max(x+y-z,x-y+z,-x+y+z,-x-y-z).
\]
The calculation shows $\Delta C=-1.233$, no evidence for contextuality. 

Search for contextuality is the specific goal of Ref. \cite{Asano},
but the meaning of the concept there is different from ours: there,
it means violations of the Suppes-Zanotti-Leggett-Garg inequality
(which is the consistently connected case of (\ref{eq:criterion n=00003D3})),
irrespective of whether these violations are due to inconsistent connectedness
or due to contextuality in our sense.

\section{Conjoint choices: Animals and sounds they make (a cyclic system of
rank 4)}

Aerts, Gabora, and Sozzo \cite{Aerts} present results of an experiment
in which each of 81 participants had to choose between two animals
and between two animal sounds, under four conditions $c_{1},c_{2},c_{3},c_{4}$
(contexts), as shown below:

\begin{center}
\begin{tabular}{cc}
 & \tabularnewline
 & \tabularnewline
\multirow{2}{*}{$V_{1}$} & \tabularnewline
 & \tabularnewline
 & \tabularnewline
\end{tabular}%
\begin{tabular}{c|c|c|c}
\multicolumn{1}{c}{} & \multicolumn{2}{c}{$W_{2}$} & \tabularnewline
\cline{2-3} 
$c_{1}$ & Growls & Whinnies & \tabularnewline
\hline 
\multicolumn{1}{|c|}{Horse} & .049 & .630 & \multicolumn{1}{c|}{.679}\tabularnewline
\hline 
\multicolumn{1}{|c|}{Bear} & .259 & .062 & \multicolumn{1}{c|}{.321}\tabularnewline
\hline 
 & .308 & .692 & \tabularnewline
\cline{2-3} 
\end{tabular}$\qquad$$\qquad$%
\begin{tabular}{c|c|c|c}
\multicolumn{1}{c}{} & \multicolumn{2}{c}{$V_{4}$} & \tabularnewline
\cline{2-3} 
$c_{4}$ & Snorts & Meows & \tabularnewline
\hline 
\multicolumn{1}{|c|}{Horse} & .593 & .025 & \multicolumn{1}{c|}{.618}\tabularnewline
\hline 
\multicolumn{1}{|c|}{Bear} & .296 & .086 & \multicolumn{1}{c|}{.382}\tabularnewline
\hline 
 & .889 & .111 & \tabularnewline
\cline{2-3} 
\end{tabular}%
\begin{tabular}{cc}
 & \tabularnewline
 & \tabularnewline
\multirow{2}{*}{$W_{1}$} & \tabularnewline
 & \tabularnewline
 & \tabularnewline
\end{tabular}
\par\end{center}

\begin{center}
\begin{tabular}{cc}
 & \tabularnewline
 & \tabularnewline
\multirow{2}{*}{$W_{3}$} & \tabularnewline
 & \tabularnewline
 & \tabularnewline
\end{tabular}%
\begin{tabular}{c|c|c|c}
\multicolumn{1}{c}{} & \multicolumn{2}{c}{$V_{2}$} & \tabularnewline
\cline{2-3} 
$c_{2}$ & Growls & Whinnies & \tabularnewline
\hline 
\multicolumn{1}{|c|}{Tiger} & .778 & .086 & \multicolumn{1}{c|}{.864}\tabularnewline
\hline 
\multicolumn{1}{|c|}{Cat} & .086 & .049 & \multicolumn{1}{c|}{.135}\tabularnewline
\hline 
 & .864 & .135 & \tabularnewline
\cline{2-3} 
\end{tabular}$\qquad$$\qquad$%
\begin{tabular}{c|c|c|c}
\multicolumn{1}{c}{} & \multicolumn{2}{c}{$W_{4}$} & \tabularnewline
\cline{2-3} 
$c_{3}$ & Snorts & Meows & \tabularnewline
\hline 
\multicolumn{1}{|c|}{Tiger} & .148 & .086 & \multicolumn{1}{c|}{.234}\tabularnewline
\hline 
\multicolumn{1}{|c|}{Cat} & .099 & .667 & \multicolumn{1}{c|}{.766}\tabularnewline
\hline 
 & .247 & .753 & \tabularnewline
\cline{2-3} 
\end{tabular}%
\begin{tabular}{cc}
 & \tabularnewline
 & \tabularnewline
\multirow{2}{*}{$V_{3}$} & \tabularnewline
 & \tabularnewline
 & \tabularnewline
\end{tabular}
\par\end{center}

\medskip{}

\noindent The ``objects'' to be measured here are the choices offered:
\[
\begin{array}{ll}
q_{1}=\textnormal{Horse or Bear?} & q_{2}=\textnormal{Growls or Whinnies?}\\
q_{3}=\textnormal{Tiger or Cat?} & q_{4}=\textnormal{Snorts or	Meows?}
\end{array}
\]
Each of the four contexts corresponds to a pair of these objects,
\[
c_{1}=(q_{1},q_{2}),c_{2}=(q_{2},q_{3}),c_{3}=(q_{3},q_{4}),c_{4}=(q_{4},q_{1}),
\]
and the choices made are binary measurements (random variables) 
\[
\begin{array}{cccc}
c_{1} & c_{2} & c_{3} & c_{4}\\
(V_{1},W_{2}) & (V_{2},W_{3}) & (V_{3},W_{4}) & (V_{4},W_{1})
\end{array}.
\]
The table of the results above translates into the diagram of expected
values

\[
\vcenter{\xymatrix@C=1cm{ & {\scriptstyle \langle V_{1}\rangle }\ar@{-}[r]_{\langle V_{1}W_{2}\rangle } & {\scriptstyle \langle W_{2}\rangle }\ar@{.}[dr]\\
{\scriptstyle \langle W_{1}\rangle }\ar@{.}[ur] &  &  & {\scriptstyle \langle V_{2}\rangle }\ar@{-}[d]_{\langle V_{2}W_{3}\rangle }\\
{\scriptstyle \langle V_{4}\rangle }\ar@{-}[u]_{\langle V_{4}W_{1}\rangle } &  &  & {\scriptstyle \langle W_{3}\rangle }\ar@{.}[dl]\\
 & {\scriptstyle \langle W_{4}\rangle }\ar@{.}[lu] & {\scriptstyle \langle V_{3}\rangle }\ar@{-}[l]_{\langle V_{3}W_{4}\rangle }
}
}=\vcenter{\xymatrix@C=1cm{ & {\scriptstyle .358}\ar@{-}[r]_{-.778} & {\scriptstyle -.384}\ar@{.}[dr]\\
{\scriptstyle .236}\ar@{.}[ur] &  &  & {\scriptstyle .728}\ar@{-}[d]_{.655}\\
{\scriptstyle .778}\ar@{-}[u]_{.358} &  &  & {\scriptstyle .728}\ar@{.}[dl]\\
 & {\scriptstyle -.506}\ar@{.}[lu] & {\scriptstyle -.532}\ar@{-}[l]_{.630}
}
}
\]
\[
\]
The noncontextuality criterion for rank 4 has the form
\begin{equation}
\Delta C=\sodd(\langle V_{1}W_{2}\rangle ,\langle V_{2}W_{3}\rangle ,\langle V_{3}W_{4}\rangle ,\langle V_{4}W_{1}\rangle )-2-\sum_{i=1}^{4}|\langle V_{i}\rangle -\langle W_{i}\rangle |\leq0,\label{eq:criterion n=00003D4}
\end{equation}
where
\[
\sodd(w,x,y,z)=\max(|w+x+y-z|,|w+x-y+z|,|w-x+y+z|,|-w+x+y+z|).
\]
The value computed from the data is $\Delta C=-3.357$, providing
no evidence for contextuality. 

Ref. \cite{Aerts} reports that contextuality in this data set is
present because
\begin{equation}
\sodd(\langle V_{1}W_{2}\rangle ,\langle V_{2}W_{3}\rangle ,\langle V_{3}W_{4}\rangle ,\langle V_{4}W_{1}\rangle )-2>0,
\end{equation}
i.e., the data violate the classical CHSH inequalities \cite{9CHSH,15Fine}.
As pointed out in Ref. \cite{DK_Topics}, the CHSH inequalities are
predicated on the assumption of consistent connectedness (marginal
selectivity). Without this assumption they cannot be derived as a
necessary or sufficient condition of noncontextuality, and this assumption
is clearly violated in the data. Aerts \cite{Aerts_response} has
developed a theory which allows for inconsistent connectedness, but
it is unclear to us how this justifies the use of CHSH inequalties
in Ref. \cite{Aerts}.

\section{\label{sec:Word-combinations:-Results}Word combinations and priming
(cyclic systems of rank 4)}

Bruza, Kitto, Ramm, and Sitbon \cite{Bruza} studied ambiguous two-word
combinations, such as ``apple chip''. One can understand this word
combination to refer to an edible chip made of apples or to an apple
computer component. It is even possible to imagine such meanings as
a piece chipped off of an apple computer, or a computer component
made of apples. In the experiments referred to the participants were
asked to explain how they understood the first and the second word
in a combination: one meaning of each word (e.g., the fruit meaning
for ``apple'', the edible product meaning for ``chip'', etc.)
can be taken for $+1$, any other meaning being classified as $-1$.
The meanings were determined by asking the participants to explain
how they understood the words. For each two-word combination the experimenters
used one of four pairs of priming words presumably affecting the meanings.
For the ``apple chip'' combination, the priming words could be
\[
\begin{array}{ll}
q_{1}=\textnormal{banana} & q_{2}=\textnormal{\textnormal{potato}}\\
q_{3}=\textnormal{\textnormal{computer}} & q_{4}=\textnormal{circuit}
\end{array},
\]
forming four contexts 
\[
\begin{array}{ll}
c_{1}=\textnormal{(banana, potato)} & c_{2}=\textnormal{\textnormal{(potato, computer)}}\\
c_{3}=\textnormal{\textnormal{(computer, circuit)}} & c_{4}=\textnormal{(circuit, banana)}
\end{array}.
\]
The order in which we list the words in a context is chosen to create
a cycle: $(q_{1},q_{2}),(q_{2},q_{3})$, etc.
Although this is not intuitive, formally, the measured ``objects''
here are the priming words $q_{1},q_{2},q_{3},q_{4}$, while the measurements
are binary random variables indicating in what meaning ($\pm1$) the
participant understood ``apple'' and ``chip''. In $(V_{1},W_{2})$
and $(V_{3},W_{4})$ the $V$'s are meanings of ``apple'' and $W$'s
the meanings of ``chip''; in $(V_{2},W_{3})$ and $(V_{4},W_{1})$
it is vice versa. (This is no more than a notational convention, purely
for the purposes of using the cyclic indexation.) 

Ref. \cite{Bruza} presents data on 23 word combinations preceded
by priming words (each combination in each context being shown to
each of 61-65 participants). In all 23 cases the computed values of
$\Delta C$ are negative, ranging from -2.882 to -0.418 (for the ``apple
chip'' example the value is -1.640). We conclude, once again, that
there is no evidence in favor of contextuality. (The authors of Ref.
\cite{Bruza} kindly provided to us the word pairs and priming words,
with the computed values of $\sodd$ and equivalents of $|\langle V_{i}\rangle -\langle W_{i}\rangle |$
($i=1,\ldots,4$), for all 23 word combinations; they are presented,
with permission, in the supplementary file S2, with the computation
of $\Delta C$ added.)

The aim of Ref. \cite{Bruza} was not to study contextuality. Rather
they were interested in the property called \emph{compositionality},
defined, in our terms, as consistent connectedness together with lack
of contextuality. Violations of this condition therefore amount to
either inconsistent connectedness or, if connectedness is consistent,
to contextuality in our sense.

\section{Psychophysical matching (cyclic systems of rank 4)}

All experiments discussed so far use participants as replicants: the
estimate of $\Pr[V=v,W=w]$ in a given context is the proportion
of participants who responded $(v,w)$, $v=\pm1$, $w=\pm1$.
In the question order effect and Schr\"oder's staircase illusion
studies different groups of people participated in different contexts,
whereas the conjoint choices and word combinations studies employed
repeated measures design: each participant made one choice in each
of the four contexts. 

In our laboratory, we searched for possible contextual effects in
a large series of psychophysical experiments where each of very few
(usually, three) participants were repeatedly ``measuring'' the
same four ``objects'' in the same four contexts. In each of the
seven experiments the number of replications per participant was 1000-2000,
evenly divided between different contexts. 

The logic of an experiment was as follows. The participant was shown
two stimuli, target one ($T$) and adjustable one ($A$), both completely
specified by two parameters. In each trial, the values $\alpha$ and
$\beta$ of these parameters (real numbers) in the target stimulus
$T(\alpha,\beta)$ are fixed at one of several values,
each pair of values determining a context; in the adjustable stimulus
the two parameters can be simultaneously or (in some experiments)
successively changed by the participant rotating a trackball. At the
end of each trial the participant reaches some values $X$ and $Y$
of these parameters that she/he judges to make $A(X,Y)$
match (i.e., look the same as) $T(\alpha,\beta)$. In most
experiments $\alpha$ and $\beta$ vary on several levels each (or
even quasi-continuously within certain ranges), and we always choose
four specific values or subranges of their values: $q_{1},q_{3}$
for $\alpha$ and $q_{2},q_{4}$ for $\beta$. They form four contexts
that can be cyclically arranged as $(q_{1},q_{2}),(q_{2},q_{3}),(q_{3},q_{4}),(q_{4},q_{1})$,
and for each of them we get empirical distributions of $X$ and $Y$:
$(X_{12},Y_{12})$ for context $(q_{1},q_{2})$,
$(X_{41},Y_{41})$ for context $(q_{4},q_{1})$,
etc. In this notation, of the two objects $q_{i},q_{j}$, the random
variable $X_{ij}$ ``measures'' the $q$ with an odd index (1 or
3), whether $i$ or $j$; analogously, $Y_{ij}$ ``measures'' the
$q$ with the even index.

The values of $X$ and $Y$ are then dichotomized in the following
way: we choose a value $x_{i}$ and a value $y_{j}$ ($i=1,3$, $j=2,4$)
and define
\begin{equation}
V_{i}=\left\{ \begin{array}{ccc}
+1 & if & X_{i,i\oplus1}>x_{i}\\
-1 & if & X_{i,i\oplus1}\leq x_{i}
\end{array}\right.,\quad V_{j}=\left\{ \begin{array}{ccc}
+1 & if & Y_{j,j\oplus1}>y_{j}\\
-1 & if & Y_{j,j\oplus1}\leq y_{j}
\end{array}\right..
\end{equation}
\begin{equation}
W_{i}=\left\{ \begin{array}{ccc}
+1 & if & X_{i\ominus1,i}>x_{i}\\
-1 & if & X_{i\ominus1,i}\leq x_{i}
\end{array}\right.,\quad W_{j}=\left\{ \begin{array}{ccc}
+1 & if & Y_{j\ominus1,j}>y_{j}\\
-1 & if & Y_{j\ominus1,j}\leq y_{j}
\end{array}\right..
\end{equation}
The values of $(x_{1},x_{3},y_{2},y_{4})$ can be chosen
in a variety of ways, and for each choice we apply to the obtained
$V$ and $W$ variables the criterion (\ref{eq:criterion n=00003D4}).

\begin{figure}
\begin{centering}
\fbox{\begin{minipage}[t]{0.45\columnwidth}%
\begin{center}
\includegraphics[bb=0bp 100bp 720bp 580bp,scale=0.25]{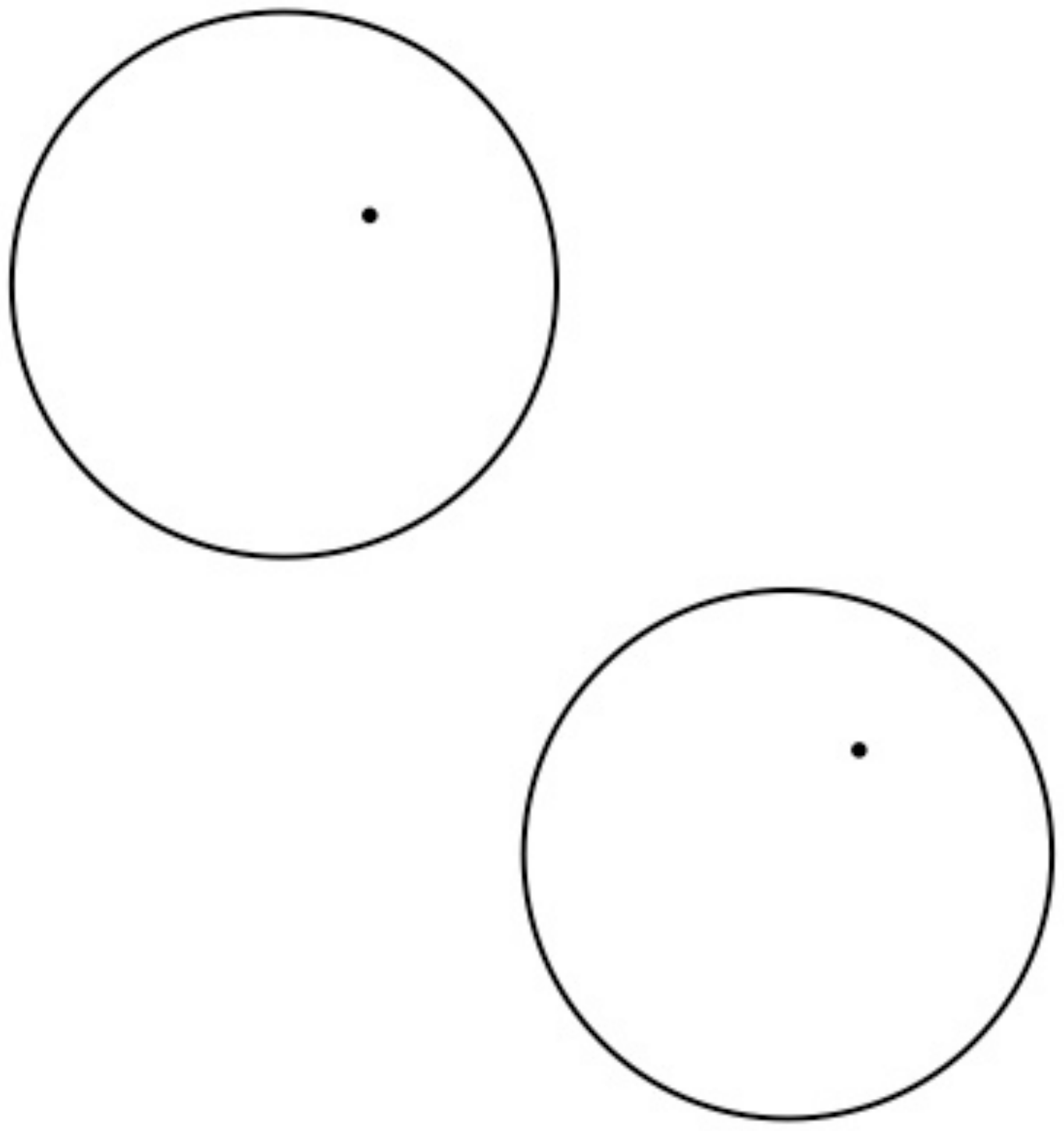}
\par\end{center}%
\end{minipage}}%
\fbox{\begin{minipage}[t]{0.45\columnwidth}%
\begin{center}
\includegraphics[bb=0bp 100bp 720bp 580bp,scale=0.25]{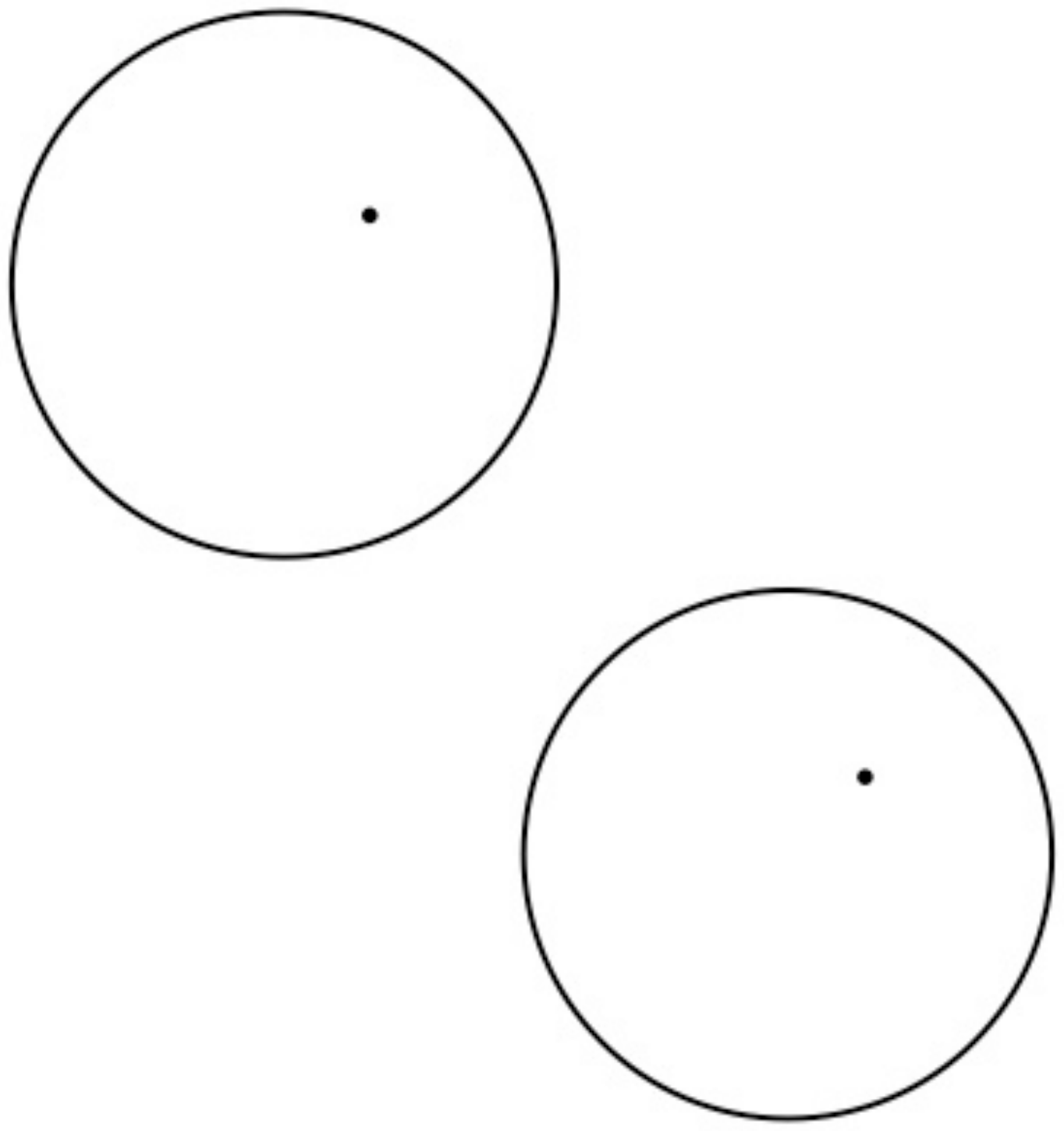}
\par\end{center}%
\end{minipage}}
\par\end{centering}

\begin{centering}
\fbox{\begin{minipage}[t]{0.45\columnwidth}%
\begin{center}
\includegraphics[bb=0bp 200bp 720bp 580bp,scale=0.25]{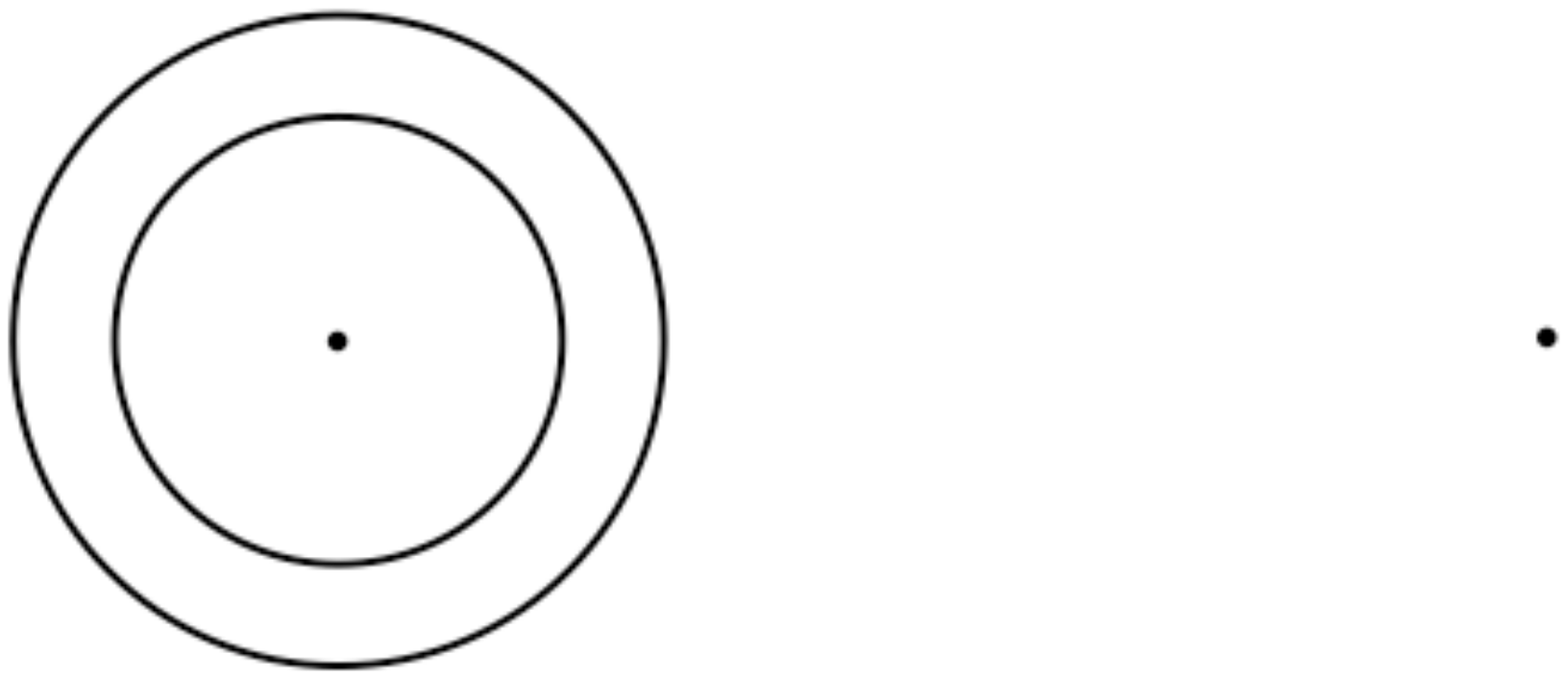}
\par\end{center}%
\end{minipage}}%
\fbox{\begin{minipage}[t]{0.45\columnwidth}%
\begin{center}
\includegraphics[bb=0bp 200bp 720bp 580bp,scale=0.25]{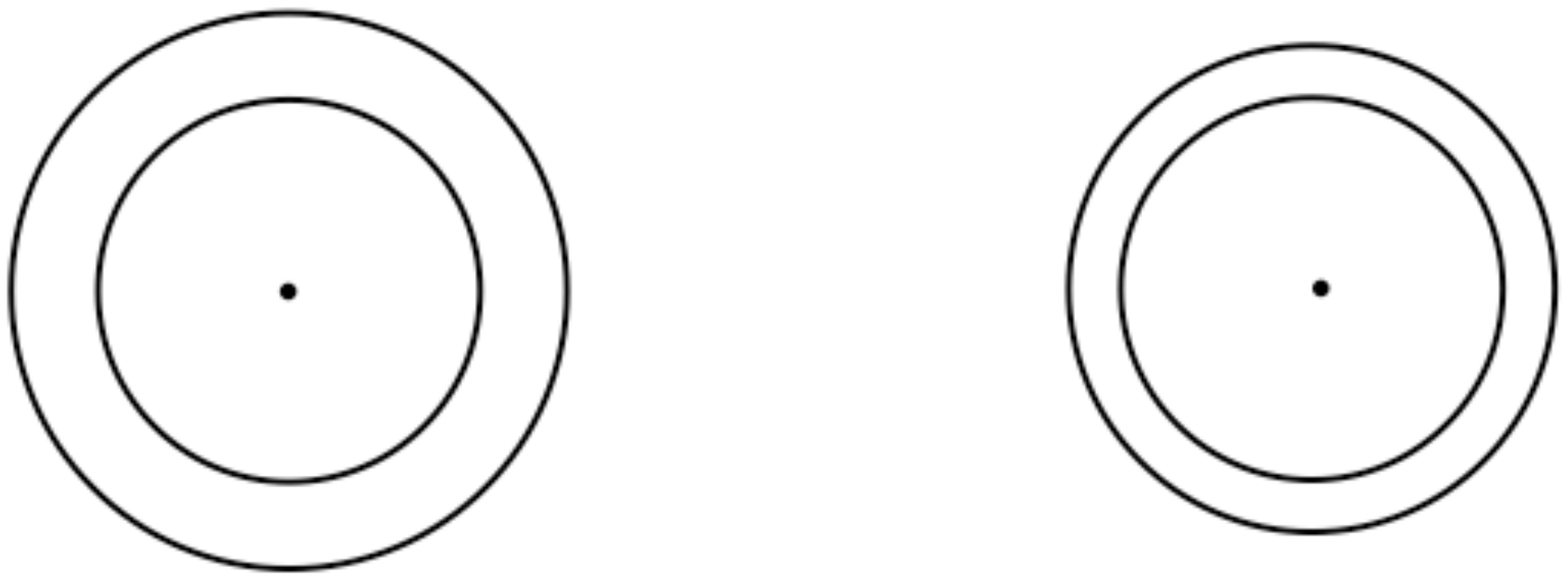}
\par\end{center}%
\end{minipage}}
\par\end{centering}

\begin{centering}
\fbox{\begin{minipage}[t]{0.45\columnwidth}%
\begin{center}
\includegraphics[bb=0bp 200bp 720bp 580bp,scale=0.25]{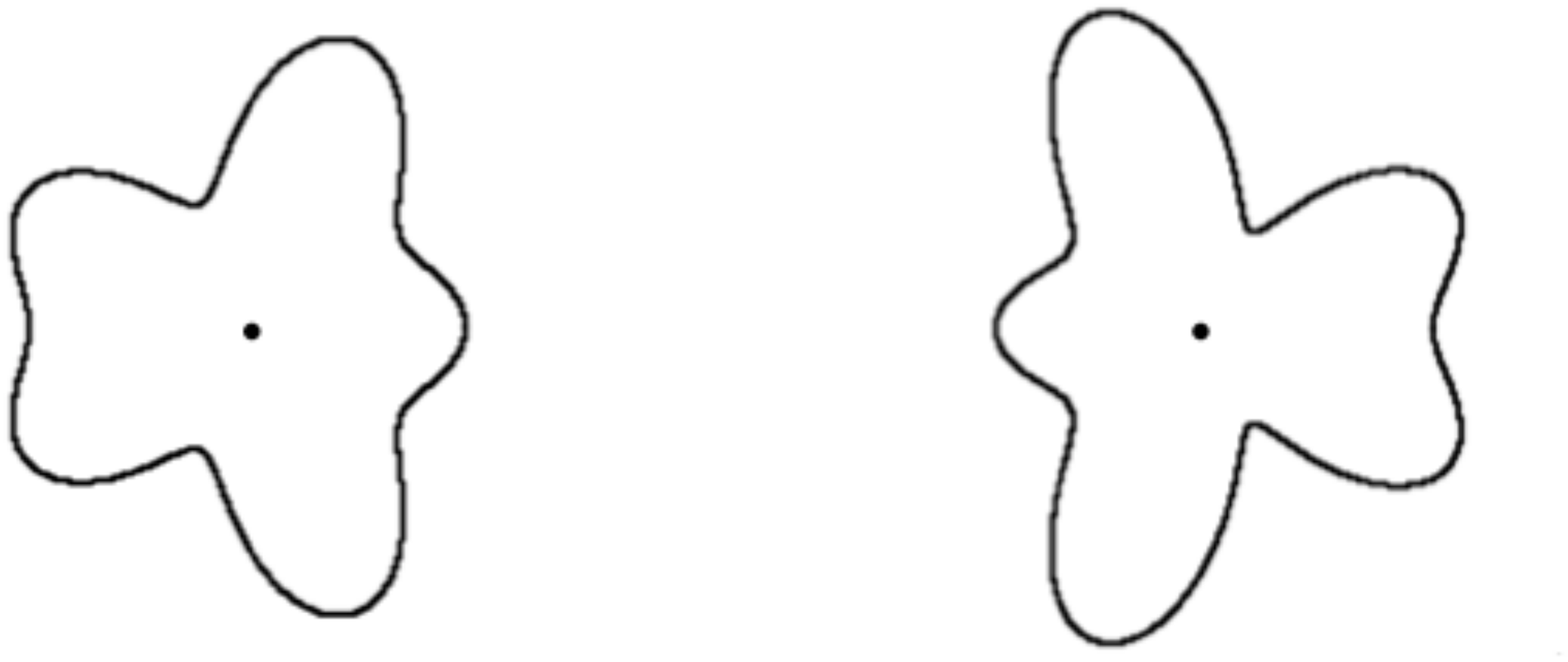}
\par\end{center}%
\end{minipage}}%
\fbox{\begin{minipage}[t]{0.45\columnwidth}%
\begin{center}
\includegraphics[bb=0bp 200bp 720bp 580bp,scale=0.25]{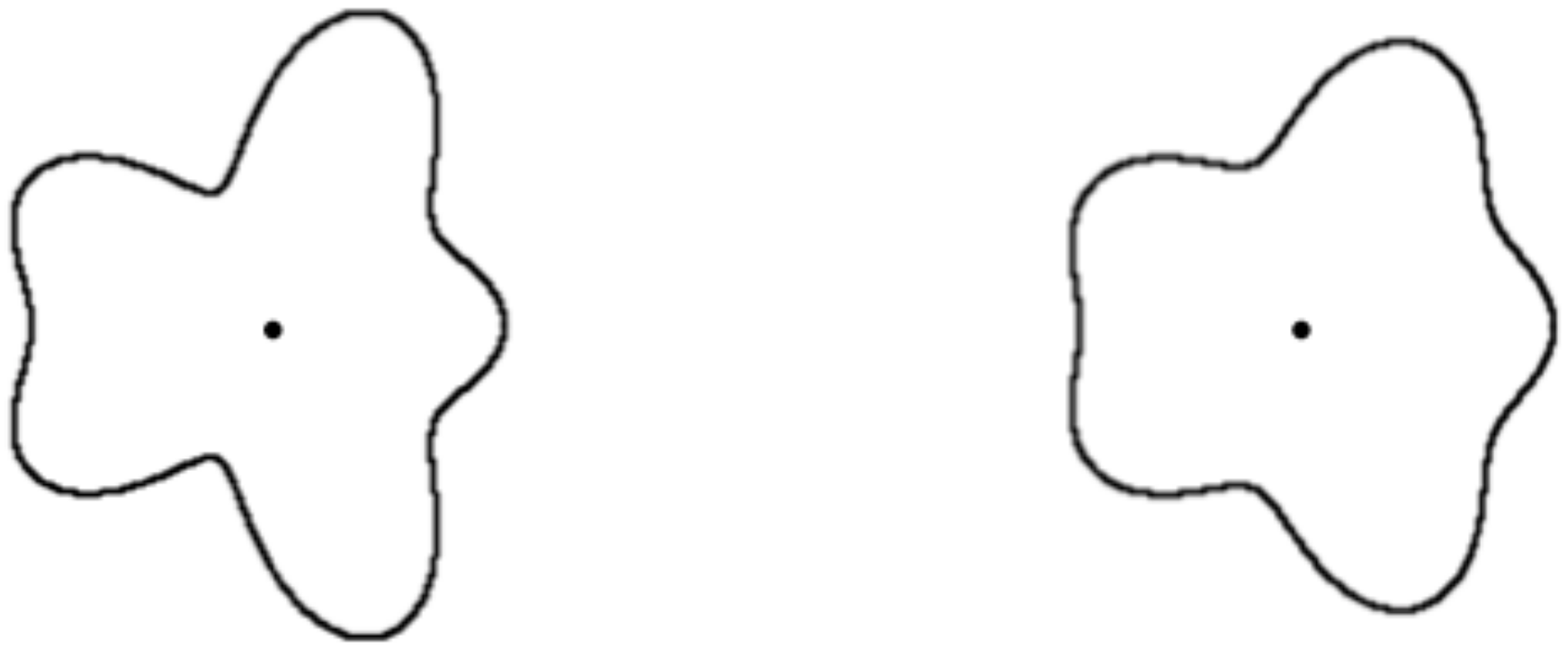}
\par\end{center}%
\end{minipage}}
\par\end{centering}

\protect\caption{Stimuli used in the matching experiments. The left panels show pairs
of stimuli at the beginning of a trial, the right panels show an intermediate
stage in the matching process. Top panels: in Experiments 1a-b there
participants adjusted the position of the dot within a lower-right
circle to match a fixed position of the target dot in the upper-left
circle. Middle panels: in Experiments 2a-c they adjusted the radii
of two concentric circles on the right to match two fixed concentric
circles on the left. Bottom panels: in Experiments 3a-b they adjusted
the amplitudes of two Fourier harmonics of a floral shape on the right
to match a fixed floral shape on the left. For details, see the supplementary
file S3.}
\end{figure}

\begin{figure}
\begin{centering}
\includegraphics[scale=0.25]{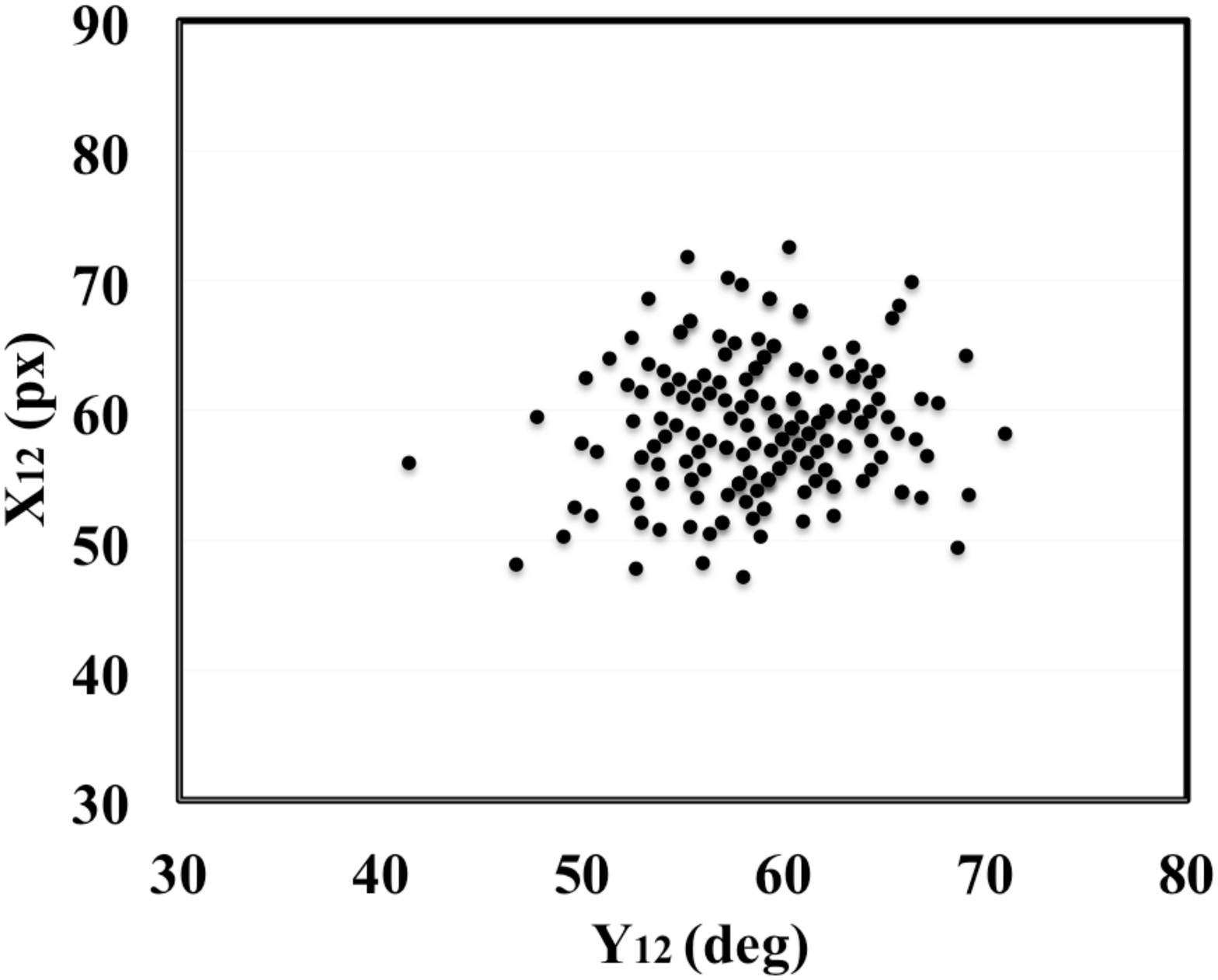}\includegraphics[scale=0.25]{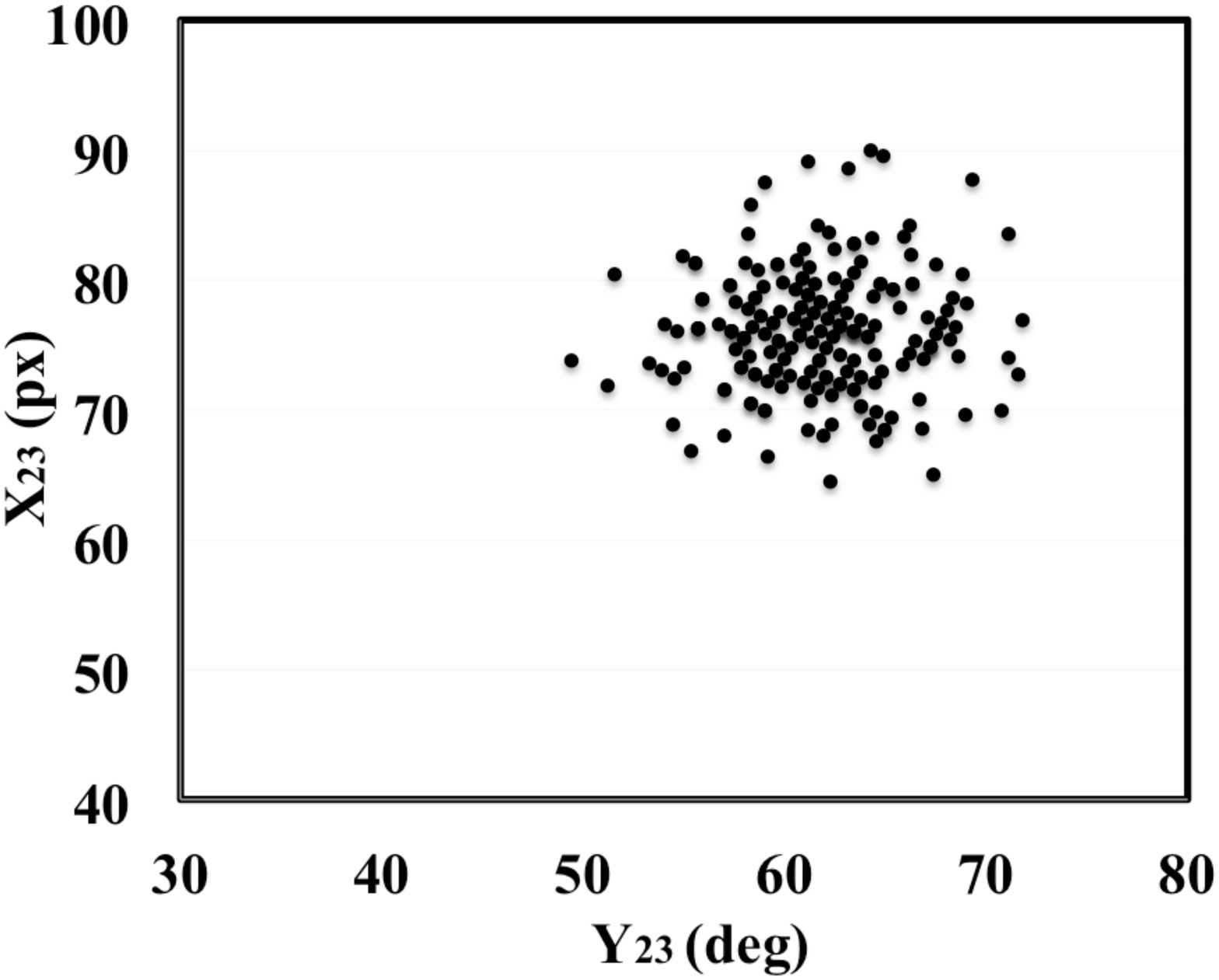}
\par\end{centering}

\begin{centering}
\includegraphics[scale=0.25]{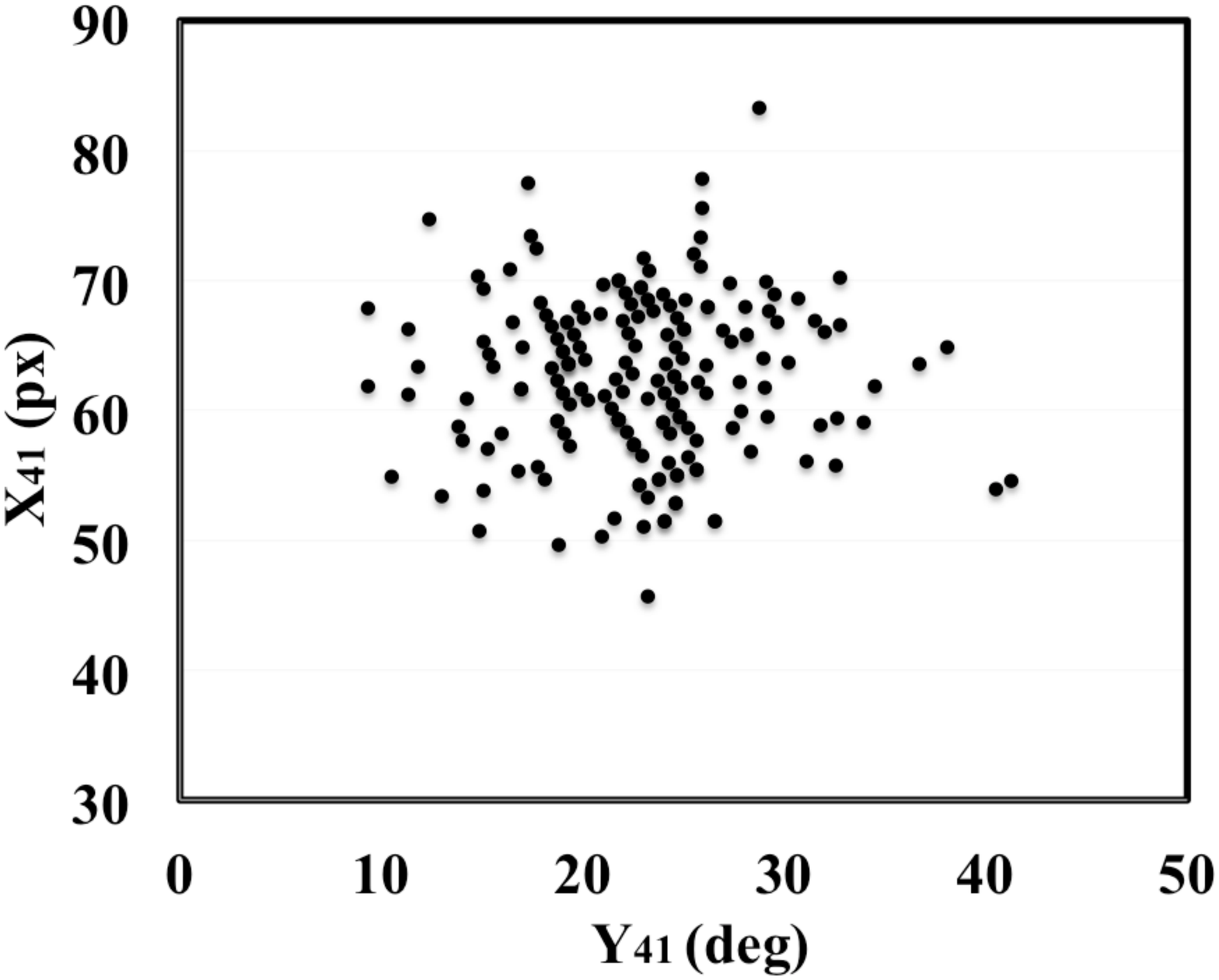}\includegraphics[scale=0.25]{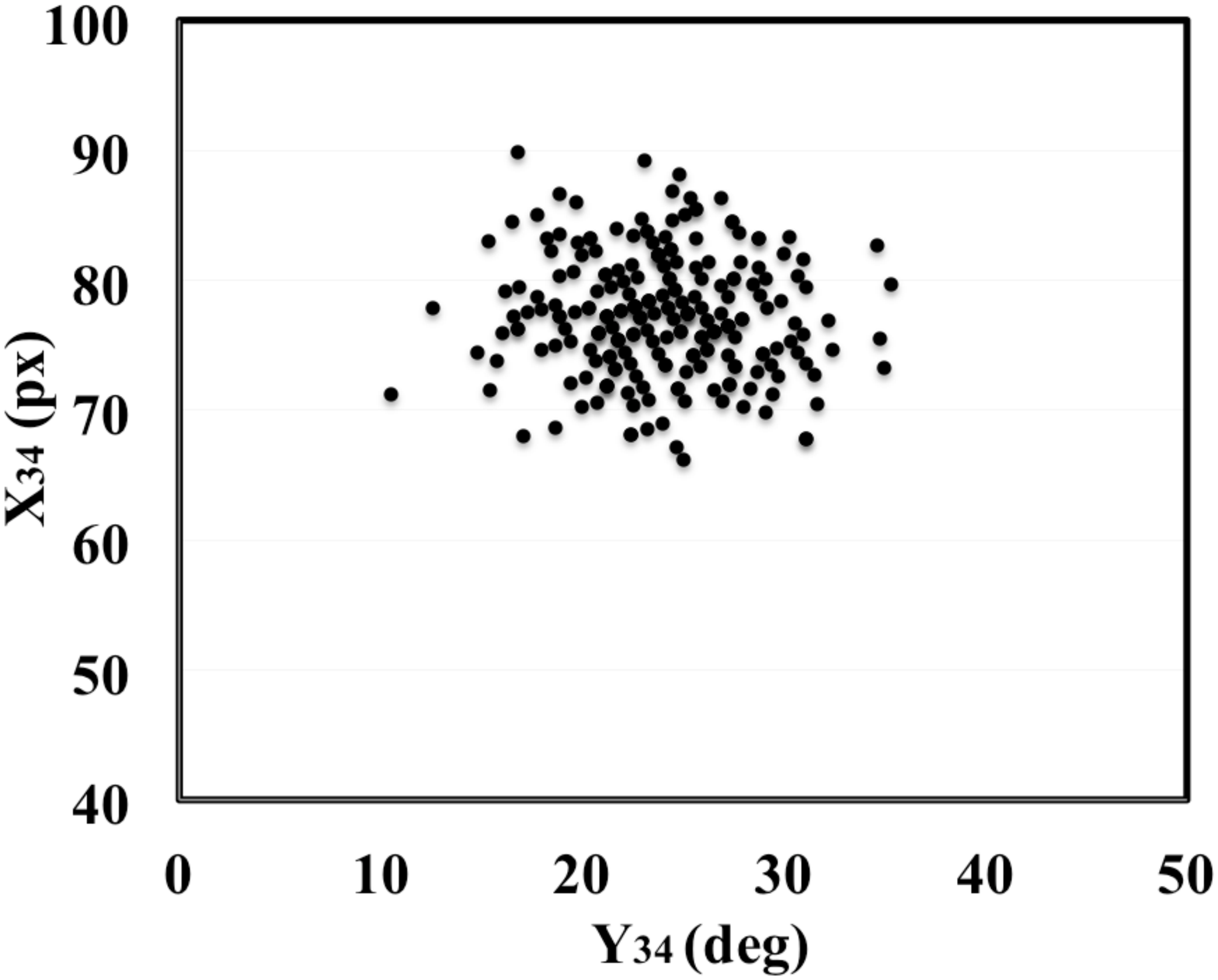}
\par\end{centering}

\protect\caption{Results for four contexts $\alpha\textnormal{ (px)}\times\beta\textnormal{ (deg)}=\{ q_{1}=53.67,q_{3}=71.55\} \times\{ q_{2}=63.43,q_{4}=26.57\} $
extracted from Experiment 1a, participant P3, about 200 replications
per context.}
\end{figure}

As an example, in one of the experiments the stimuli $T$ and $A$
were two dots in two circles, like the ones shown in Figure 2, top,
with a dot's position within a circle described in polar coordinates
($\alpha$ and $X$ denoting distance from the center in pixels, $\beta$
and $Y$ denoting angle in degrees measured counterclockwise from
the horizontal rightward radius-vector). We extract from this experiment
a $2\times2$ subdesign as shown in Figure 3. Then we choose a value
of $x_{1}$ as any integer (in pixels) between $\mathrm{max}[\mathrm{min}X_{12},\mathrm{min}X_{41}]$
and $\mathrm{min}[\mathrm{max}X_{12},\mathrm{max}X_{41}]$, we choose
$y_{2}$ as any integer (in degrees) between $\mathrm{max}[\mathrm{min}Y_{12},\mathrm{min}Y_{23}]$
and $\mathrm{min}[\mathrm{max}Y_{12},\mathrm{max}Y_{23}]$, and analogously
for $x_{3}$ and $y_{4}$. This yields $25\times23\times21\times79$
quadruples of $(x_{1},x_{3},y_{2},y_{4})$, and the corresponding
number of cyclic systems of binary random variables $(V_{1},W_{2},V_{2},W_{3},V_{3},W_{4},V_{4},W_{1})$.
Consider, e.g., one such choice: $(x_{1},x_{3},y_{2},y_{4})=(72\textnormal{ px},67\textnormal{ px},60\textnormal{ deg},23\textnormal{ deg})$.
The diagram of this system is
\[
\vcenter{\xymatrix@C=1cm{ & {\scriptstyle \langle V_{1}\rangle }\ar@{-}[r]_{\langle V_{1}W_{2}\rangle } & {\scriptstyle \langle W_{2}\rangle }\ar@{.}[dr]\\
{\scriptstyle \langle W_{1}\rangle }\ar@{.}[ur] &  &  & {\scriptstyle \langle V_{2}\rangle }\ar@{-}[d]_{\langle V_{2}W_{3}\rangle }\\
{\scriptstyle \langle V_{4}\rangle }\ar@{-}[u]_{\langle V_{4}W_{1}\rangle } &  &  & {\scriptstyle \langle W_{3}\rangle }\ar@{.}[dl]\\
 & {\scriptstyle \langle W_{4}\rangle }\ar@{.}[lu] & {\scriptstyle \langle V_{3}\rangle }\ar@{-}[l]_{\langle V_{3}W_{4}\rangle }
}
}=\vcenter{\xymatrix@C=1cm{ & {\scriptstyle -.989}\ar@{-}[r]_{.211} & {\scriptstyle -0.2}\ar@{.}[dr]\\
{\scriptstyle -0.902}\ar@{.}[ur] &  &  & {\scriptstyle 0.300}\ar@{-}[d]_{.301}\\
{\scriptstyle -0.006}\ar@{-}[u]_{.016} &  &  & {\scriptstyle 0.960}\ar@{.}[dl]\\
 & {\scriptstyle 0.167}\ar@{.}[lu] & {\scriptstyle 0.991}\ar@{-}[l]_{.158}
}
}
\]
\[
\]
and the value of $\Delta C=-2.137$, no evidence of contextuality.
In fact negative values of $\Delta C$ are obtained for all $25\times23\times21\times79$
dichotomizations. Clearly, different dichotomizations of the same
random variables are not stochastically independent, but there is
no mathematical reason for $\Delta C$ to be of the same sign in all
of them. 

In the supplementary file S3 we describe in detail how the dichotomizations
were made, their number ranging from 3024 to 11,663,568 per $2\times2$
(sub)design in each experiment for each participant. The outcome is:
not a single case with positive $\Delta C$ observed.

\section{Conclusion}

The empirical data analyzed above suggest that the noncontextuality
boundaries, that are generally breached in quantum physics, are not
breached by behavioral and social systems. This may seem a disappointing
conclusion for some. With the realization that quantum formalisms
may be used to construct models in various areas outside physics \cite{BB-book,HavenKhrennikov,Khrennikov2010,OHYaVolovich2011},
the expectation was created that human behavior should exhibit contextuality,
perhaps even on a greater scale than allowed by quantum theory. However,
if the no-contextuality conclusion of the present paper is proved
to be a rule for a very broad class of behavioral and social systems, it is rather fortunate for behavioral and social
sciences. Noncontextuality means more constrained behavior, and constraints
allow one to make predictions. The power of quantum mechanics is not
in that quantum systems breach the classical-mechanical bounds of
noncontextuality, but in the theory that imposes other, equally strict
constraints. Presence of contextuality, in the absence of a general
theory like quantum mechanics, translates into unpredictability.

It must be noted that absence of contextuality in behavioral and social
systems does not mean that quantum formalisms are not applicable to
them. A good argument for why this conclusion would be groundless
is provided by the question order effect discussed in Section \ref{sec:Question-order-effect}:
it is precisely the applicability of a quantum-mechanical model in
the question order effect analysis \cite{Wang-Busemeyer,Wang} that
allows one to predict the lack of contextuality in this paradigm.

When discussing contextuality, one should be aware of the likelihood
of purely terminological confusions. It is clear that in the behavioral
and social systems a context generally influences the measurement
of an object within it. For instance, the distribution of answers
to a question depends on a question asked and answered before it.
One could call this contextuality, and many do. This is, however,
a trivial sense of contextuality, on a par with the fact that the
distribution of answers to a question depends on what this question
is. One should not be surprised that other factors (such as temperature
in the lab or questions asked and answered previously) can influence
this distribution too. We call this inconsistent connectedness, and
we offer a theory that distinguishes this ubiquitous feature from
contextuality in a different, one could argue more interesting meaning.

\subsection*{Acknowledgments}
This research has been supported by NSF grant SES-1155956, AFOSR grant FA9550-14-1-0318, and A. von Humboldt Foundation. We are grateful to the authors of Refs.  \cite{Asano}, \cite{Bruza}, and \cite{Wang} for kindly providing data sets for our analysis. We have benefited from discussions with
Jan-{\AA}ke Larsson and Victor H. Cervantes (who pointed out a mistake in an earlier version of the paper). The computations discussed in Sections 3 and 6 are presented in the supplementary files S1 and S2, respectively. The original data sets are available from the authors of Refs. \cite{Bruza} and  \cite{Wang}. Details of the experiments discussed in Section 7 are presented in the supplementary file S3; the data sets are available as "Contextuality in Psychophysical Matching", http://dx.doi.org/10.7910/DVN/OJZKKP, Harvard Dataverse, V1.




\section*{Supplementary material}

| Resp to A | | Resp to B | | | | | | | | | | | | |
|---|---|---|---|---|---|---|---|---|---|---|---|---|---|---|
| order A -> B | order B -> A | order A -> B | order B -> A | | | | | | | | | | | |
| <V1> | <W1> | <W2> | <V2> | <V1*W2> | <V2*W1> | s1-value | \|<V1>-<W1>\|+\|<V2>-<W2>\| | Delta C | Delta C > 0? | QQ chi^2 | QQ p<0.05? | QQ p<0.01? | Data set # | |
| 0.2105 | 0.0482 | -0.1100 | -0.1491 | 0.5359 | 0.4781 | 0.0578 | 0.2014 | -0.1435 | 0 | 0.9830 | 0 | 0 | 1 | PEW |
| 0.2584 | 0.2081 | 0.0096 | 0.0407 | 0.5694 | 0.5882 | 0.0189 | 0.0814 | -0.0625 | 0 | 0.1150 | 0 | 0 | 2 | PEW |
| 0.3398 | 0.2705 | 0.1359 | 0.0984 | 0.6214 | 0.5820 | 0.0394 | 0.1069 | -0.0675 | 0 | 0.2736 | 0 | 0 | 3 | PEW |
| -0.0484 | -0.0324 | -0.2823 | -0.2146 | 0.5806 | 0.5749 | 0.0057 | 0.0837 | -0.0779 | 0 | 0.0123 | 0 | 0 | 4 | PEW |
| 0.1239 | 0.1367 | -0.1193 | 0.0325 | 0.6193 | 0.6529 | 0.0337 | 0.1646 | -0.1309 | 0 | 0.4268 | 0 | 0 | 5 | PEW |
| 0.0136 | 0.1198 | -0.1900 | -0.0645 | 0.6606 | 0.6406 | 0.0201 | 0.2318 | -0.2117 | 0 | 0.1531 | 0 | 0 | 6 | PEW |
| 0.0411 | 0.0206 | -0.2055 | -0.1512 | 0.6370 | 0.5945 | 0.0425 | 0.0748 | -0.0323 | 0 | 0.8474 | 0 | 0 | 7 | PEW |
| -0.1442 | -0.0946 | -0.3759 | -0.2634 | 0.5981 | 0.6675 | 0.0694 | 0.1620 | -0.0926 | 0 | 1.6353 | 0 | 0 | 8 | PEW |
| -0.0494 | -0.0847 | -0.2565 | -0.2037 | 0.6424 | 0.6156 | 0.0268 | 0.0881 | -0.0613 | 0 | 0.2559 | 0 | 0 | 9 | PEW |
| -0.0432 | -0.1231 | -0.2815 | -0.3588 | 0.6304 | 0.5633 | 0.0671 | 0.1571 | -0.0900 | 0 | 2.0235 | 0 | 0 | 10 | PEW |
| -0.2070 | -0.1544 | -0.3815 | -0.2354 | 0.6259 | 0.5139 | 0.1120 | 0.1987 | -0.0866 | 0 | 3.6979 | 0 | 0 | 11 | PEW |
| 0.3515 | 0.3293 | -0.4387 | -0.4622 | 0.1662 | 0.1118 | 0.0544 | 0.0457 | 0.0087 | 1 | 0.5272 | 0 | 0 | 12 | PEW |
| 0.3099 | 0.1750 | -0.3568 | -0.3650 | 0.2770 | 0.3700 | 0.0930 | 0.1431 | -0.0500 | 0 | 1.9950 | 0 | 0 | 13 | PEW |
| 0.1979 | 0.2245 | -0.3298 | -0.2857 | 0.3562 | 0.3878 | 0.0316 | 0.0707 | -0.0391 | 0 | 0.2227 | 0 | 0 | 14 | PEW |
| 0.1715 | 0.0633 | -0.3487 | -0.3321 | 0.3950 | 0.4894 | 0.0945 | 0.1248 | -0.0304 | 0 | 2.8834 | 0 | 0 | 15 | PEW |
| 0.1789 | 0.0691 | -0.4252 | -0.4362 | 0.3372 | 0.3351 | 0.0021 | 0.1207 | -0.1185 | 0 | 0.0009 | 0 | 0 | 16 | PEW |
| 0.0651 | 0.0815 | -0.4943 | -0.3718 | 0.3870 | 0.4115 | 0.0246 | 0.1389 | -0.1143 | 0 | 0.1838 | 0 | 0 | 17 | PEW |
| 0.0805 | -0.0056 | -0.5862 | -0.5562 | 0.3103 | 0.2921 | 0.0182 | 0.1161 | -0.0979 | 0 | 0.0642 | 0 | 0 | 18 | PEW |
| 0.0291 | 0.0132 | -0.5058 | -0.5145 | 0.3721 | 0.3140 | 0.0581 | 0.0246 | 0.0335 | 1 | 0.6917 | 0 | 0 | 19 | PEW |
| 0.1358 | -0.0751 | -0.2741 | -0.3888 | 0.5012 | 0.4621 | 0.0391 | 0.2676 | -0.2285 | 0 | 0.4058 | 0 | 0 | 20 | PEW |
| 0.0074 | -0.0076 | -0.4604 | -0.4582 | 0.6042 | 0.4025 | 0.0603 | 0.0172 | 0.0432 | 1 | 1.7896 | 0 | 0 | 21 | PEW |
| 0.0030 | -0.0877 | -0.5137 | -0.4854 | 0.4225 | 0.3567 | 0.0658 | 0.1191 | -0.0533 | 0 | 0.8554 | 0 | 0 | 22 | PEW |
| -0.0620 | -0.0392 | -0.5682 | -0.5196 | 0.4342 | 0.3107 | 0.1235 | 0.0715 | 0.0520 | 1 | 3.4820 | 0 | 0 | 23 | PEW |
| 0.0278 | 0.0997 | -0.4889 | -0.4394 | 0.4056 | 0.3100 | 0.0956 | 0.1215 | -0.0259 | 0 | 1.9147 | 0 | 0 | 24 | PEW |
| 0.0290 | -0.0261 | -0.5411 | -0.4774 | 0.3816 | 0.3682 | 0.0135 | 0.1187 | -0.1053 | 0 | 0.0441 | 0 | 0 | 25 | PEW |
| 0.0230 | -0.0300 | 0.0138 | -0.0700 | -0.4286 | -0.3000 | 0.1286 | 0.1369 | -0.0083 | 0 | 1.9870 | 0 | 0 | 26 | PEW |
| -0.0876 | -0.0729 | -0.0046 | -0.0729 | -0.2258 | -0.1667 | 0.0591 | 0.0829 | -0.0238 | 0 | 0.3719 | 0 | 0 | 27 | PEW |
| 0.0486 | -0.0960 | -0.1243 | -0.1818 | -0.2000 | -0.1263 | 0.0737 | 0.2021 | -0.1284 | 0 | 1.0697 | 0 | 0 | 28 | PEW |
| -0.0611 | -0.1761 | -0.1444 | -0.2159 | 0.0056 | -0.0398 | 0.0453 | 0.1865 | -0.1412 | 0 | 0.3658 | 0 | 0 | 29 | PEW |
| -0.1940 | -0.1705 | -0.2239 | -0.1705 | 0.1343 | 0.0650 | 0.0693 | 0.0770 | -0.0077 | 0 | 0.9938 | 0 | 0 | 30 | PEW |
| -0.1279 | -0.2044 | -0.0444 | -0.1124 | -0.0548 | -0.1182 | 0.0633 | 0.1447 | -0.0814 | 0 | 0.7374 | 0 | 0 | 31 | PEW |
| -0.2185 | -0.3063 | -0.2185 | -0.1848 | 0.0231 | -0.0228 | 0.0459 | 0.1215 | -0.0756 | 0 | 0.4133 | 0 | 0 | 32 | PEW |
| -0.2204 | -0.4018 | 0.0863 | 0.1239 | -0.1757 | -0.1178 | 0.0579 | 0.2190 | -0.1611 | 0 | 0.5515 | 0 | 0 | 33 | PEW |
| -0.3199 | -0.3651 | -0.0428 | -0.0370 | -0.1261 | -0.0794 | 0.0365 | 0.0510 | -0.0144 | 0 | 0.2597 | 0 | 0 | 34 | PEW |
| -0.4142 | -0.3803 | -0.1420 | -0.2169 | 0.0414 | 0.0817 | 0.0403 | 0.1088 | -0.0685 | 0 | 0.2820 | 0 | 0 | 35 | PEW |
| -0.2845 | -0.2044 | -0.2044 | -0.1261 | -0.0899 | 0.0803 | 0.0362 | 0.2581 | -0.2219 | 0 | 0.2344 | 0 | 0 | 36 | PEW |
| -0.3433 | -0.3008 | -0.1164 | -0.0786 | -0.0388 | 0.0081 | 0.0469 | 0.0803 | -0.0334 | 0 | 0.3878 | 0 | 0 | 37 | PEW |
| -0.3598 | -0.4525 | -0.2691 | -0.1844 | 0.0255 | -0.0503 | 0.0758 | 0.1775 | -0.1017 | 0 | 1.0208 | 0 | 0 | 38 | PEW |
| -0.2978 | -0.2213 | -0.2089 | -0.1762 | -0.1022 | -0.0451 | 0.0571 | 0.1091 | -0.0520 | 0 | 0.7698 | 0 | 0 | 39 | PEW |
| -0.4000 | -0.3265 | -0.3118 | -0.1877 | -0.0526 | -0.1465 | 0.0939 | 0.2017 | -0.1078 | 0 | 1.7120 | 0 | 0 | 40 | PEW |
| -0.4677 | -0.4707 | -0.4160 | -0.2977 | 0.0801 | -0.0382 | 0.1182 | 0.1213 | -0.0031 | 0 | 2.7289 | 0 | 0 | 41 | PEW |
| 0.0576 | 0.0335 | -0.0262 | 0.1196 | 0.6859 | 0.6842 | 0.0017 | 0.1699 | -0.1682 | 0 | 0.0005 | 0 | 0 | 42 | PEW |
| -0.0343 | 0.0554 | -0.0218 | 0.0321 | 0.7632 | 0.7901 | 0.0268 | 0.1435 | -0.1167 | 0 | 0.3015 | 0 | 0 | 43 | PEW |
| -0.1736 | -0.1304 | -0.1240 | -0.1237 | 0.7741 | 0.7659 | 0.0082 | 0.0654 | -0.0572 | 0 | 0.0275 | 0 | 0 | 44 | PEW |
| 0.0360 | -0.0455 | -0.4676 | -0.5649 | -0.2230 | -0.2208 | 0.0022 | 0.1787 | -0.1765 | 0 | 0.0008 | 0 | 0 | 45 | PEW |
| 0.4067 | 0.3800 | -0.2575 | -0.0019 | 0.0089 | 0.0888 | 0.0800 | 0.2824 | -0.2025 | 0 | 1.7499 | 0 | 0 | 46 | PEW |
| 0.0846 | 0.0777 | 0.1741 | 0.3368 | 0.6318 | 0.5337 | 0.0982 | 0.1695 | -0.0714 | 0 | 1.4372 | 0 | 0 | 47 | PEW |
| 0.1481 | 0.1743 | -0.0617 | 0.0642 | 0.6914 | 0.7798 | 0.0885 | 0.1521 | -0.0637 | 0 | 1.6198 | 0 | 0 | 48 | PEW |
| 0.3838 | 0.2394 | 0.1616 | 0.1737 | 0.6768 | 0.6826 | 0.1448 | 0.1565 | -0.0117 | 0 | 4.9124 | 1 | 0 | 49 | PEW |
| -0.5987 | -0.6622 | -0.1973 | -0.1824 | 0.3043 | 0.2500 | 0.0543 | 0.0784 | -0.0240 | 0 | 0.4759 | 0 | 0 | 50 | PEW |
| -0.0236 | -0.1360 | -0.0866 | -0.2960 | 0.5906 | 0.6160 | 0.0254 | 0.3218 | -0.2963 | 0 | 0.0441 | 0 | 0 | 51 | PEW |
| -0.3111 | -0.4415 | -0.4833 | -0.5847 | 0.5278 | 0.5990 | 0.0713 | 0.2318 | -0.1605 | 0 | 1.4492 | 0 | 0 | 52 | PEW |
| 0.2000 | 0.3135 | -0.0173 | 0.0618 | -0.1464 | -0.1304 | 0.0350 | 0.1926 | -0.1576 | 0 | 0.2636 | 0 | 0 | 53 | PEW |
| 0.5371 | 0.5601 | 0.2800 | 0.3314 | 0.0686 | 0.0557 | 0.0129 | 0.0744 | -0.0615 | 0 | 0.0286 | 0 | 0 | 54 | PEW |
| -0.4947 | -0.5130 | -0.5211 | -0.4404 | 0.2368 | 0.1917 | 0.0451 | 0.0989 | -0.0537 | 0 | 0.4088 | 0 | 0 | 55 | PEW |
| 0.0684 | 0.0957 | -0.0769 | -0.0522 | 0.6325 | 0.6957 | 0.0632 | 0.0520 | 0.0111 | 1 | 0.8281 | 0 | 0 | 56 | PEW |
| 0.2874 | 0.3117 | -0.1494 | -0.0931 | 0.3487 | 0.3522 | 0.0036 | 0.0807 | -0.0771 | 0 | 0.0018 | 0 | 0 | 57 | PEW |
| -0.8069 | -0.6839 | -0.8414 | -0.7871 | 0.7310 | 0.6129 | 0.1181 | 0.1773 | -0.0592 | 0 | 3.8166 | 0 | 0 | 58 | PEW |
| 0.3173 | 0.3548 | 0.0481 | -0.0215 | 0.2885 | 0.3656 | 0.0771 | 0.1071 | -0.0300 | 0 | 0.6562 | 0 | 0 | 59 | PEW |
| -0.6582 | -0.5960 | 0.0408 | 0.1010 | -0.1276 | -0.3131 | 0.1856 | 0.1224 | 0.0632 | 1 | 7.1309 | 1 | 1 | 60 | PEW |
| -0.1048 | -0.1416 | -0.4531 | -0.3307 | 0.3629 | 0.3622 | 0.0007 | 0.3439 | -0.3432 | 0 | 0.0001 | 0 | 0 | 61 | PEW |
| -0.0797 | -0.0842 | 0.3768 | 0.5421 | -0.2536 | -0.2997 | 0.0460 | 0.1697 | -0.1237 | 0 | 0.3288 | 0 | 0 | 62 | PEW |
| 0.2600 | 0.1906 | -0.0680 | -0.0650 | 0.4822 | 0.5291 | 0.0469 | 0.0723 | -0.0254 | 0 | 1.3466 | 0 | 0 | 63 | PEW |
| -0.1441 | -0.0091 | -0.7205 | -0.7455 | 0.3013 | 0.2273 | 0.0740 | 0.1599 | -0.0859 | 0 | 0.6615 | 0 | 0 | 64 | PEW |
| -0.3260 | -0.2665 | -0.0313 | -0.0621 | 0.0645 | 0.0421 | 0.0224 | 0.0903 | -0.0679 | 0 | 0.1308 | 0 | 0 | 65 | PEW |
| -0.2343 | -0.3559 | 0.1033 | -0.0056 | -0.0730 | -0.0066 | 0.0674 | 0.2306 | -0.1632 | 0 | 0.8542 | 0 | 0 | 66 | PEW |
| 0.0694 | 0.1759 | 0.3333 | 0.5231 | 0.5570 | 0.5509 | 0.0061 | 0.2964 | -0.2903 | 0 | 0.0119 | 0 | 0 | 67 | Gallup |
| -0.0481 | -0.1830 | 0.4279 | 0.3740 | 0.1945 | 0.1883 | 0.0062 | 0.1889 | -0.1827 | 0 | 0.0081 | 0 | 0 | 68 | Gallup |
| -0.1678 | 0.0782 | 0.1198 | -0.0782 | 0.6427 | 0.6049 | 0.0378 | 0.4440 | -0.4062 | 0 | 0.5515 | 0 | 0 | 69 | Gallup |
| -0.3793 | 0.0741 | 0.0862 | -0.2222 | 0.2931 | 0.4444 | 0.1513 | 0.7618 | -0.6105 | 0 | 1.4845 | 0 | 0 | 70 | see PNAS 2014 paper |
| 0.0508 | 0.1887 | 0.2881 | 0.3962 | 0.5932 | 0.6792 | 0.0860 | 0.2459 | -0.1599 | 0 | 0.6963 | 0 | 0 | 71 | see PNAS 2014 paper |
| 0.6792 | 0.2131 | -0.0375 | 0.6590 | 0.2423 | 0.4230 | 0.1806 | 1.1626 | -0.9820 | 0 | 5.4846 | 1 | 0 | 72 | see PNAS 2014 paper |
| 0.4111 | 0.5606 | 0.6166 | 0.3333 | 0.1680 | 0.1537 | 0.0143 | 0.8656 | -0.8513 | 0 | 25.4164 | 1 | 1 | 73 | Gallup: Rose-Jackson |

| word combination | prime q1 | prime q3 | prime q2 | prime q4 | s1-value | \|<V1>-<W1>\| | \|<V2>-<W2>\| | \|<V3>-<W3>\| | \|<V4>-<W4>\| | N | Delta C |
|---|---|---|---|---|---|---|---|---|---|---|---|
| boxer bat | dog | fighter | ball | vampire | 0.740 | 0.350 | 0.676 | 0.280 | 0.316 | 64.000 | -2.882 |
| table file | chair | chart | nail | folder | 0.330 | 0.116 | 0.228 | 0.470 | 0.226 | 63.000 | -2.710 |
| star suit | moon | movie | vest | law | 1.180 | 0.616 | 0.108 | 0.326 | 0.116 | 62.000 | -1.986 |
| mole pen | dig | face | pig | ink | 1.180 | 0.250 | 0.126 | 0.042 | 0.600 | 63.000 | -1.838 |
| crane hatch | lift | bird | door | egg | 1.920 | 0.282 | 0.298 | 0.592 | 0.466 | 63.000 | -1.718 |
| stag yarn | party | deer | story | wool | 1.770 | 0.750 | 0.208 | 0.438 | 0.090 | 61.000 | -1.716 |
| apple chip | banana | computer | potato | circuit | 2.110 | 0.500 | 0.588 | 0.228 | 0.434 | 65.000 | -1.640 |
| web bug | spider | internet | beetle | computer | 2.000 | 0.420 | 0.592 | 0.134 | 0.306 | 63.000 | -1.452 |
| bank log | money | river | journal | tree | 2.130 | 0.110 | 0.676 | 0.184 | 0.514 | 65.000 | -1.354 |
| port vessel | harbour | wine | ship | bottle | 1.560 | 0.212 | 0.226 | 0.170 | 0.236 | 65.000 | -1.284 |
| count watch | number | dracula | time | look | 1.400 | 0.390 | 0.022 | 0.126 | 0.126 | 65.000 | -1.264 |
| rock strike | stone | music | hit | union | 2.010 | 0.376 | 0.626 | 0.234 | 0.026 | 64.000 | -1.252 |
| fan post | football | cool | mail | light | 2.130 | 0.700 | 0.050 | 0.250 | 0.376 | 63.000 | -1.246 |
| match bowl | flame | contest | disk | throw | 1.750 | 0.274 | 0.150 | 0.500 | 0.044 | 64.000 | -1.218 |
| seal pack | walrus | envelop | leader | suitcase | 2.140 | 0.166 | 0.324 | 0.426 | 0.442 | 64.000 | -1.218 |
| spring plant | summer | coil | leaf | factory | 2.020 | 0.588 | 0.000 | 0.266 | 0.346 | 64.000 | -1.180 |
| slug duck | snail | punch | quack | dodge | 1.830 | 0.192 | 0.266 | 0.306 | 0.052 | 63.000 | -0.986 |
| bill scale | phone | pelican | weight | fish | 1.630 | 0.162 | 0.108 | 0.226 | 0.108 | 64.000 | -0.974 |
| toast gag | jam | speech | choke | joke | 1.230 | 0.000 | 0.036 | 0.016 | 0.052 | 63.000 | -0.874 |
| net cap | gain | volleyball | limit | hat | 1.860 | 0.070 | 0.118 | 0.184 | 0.350 | 65.000 | -0.862 |
| battery charge | car | assault | volt | prosecute | 2.010 | 0.134 | 0.234 | 0.096 | 0.240 | 63.000 | -0.694 |
| club bar | member | golf | pub | handle | 2.280 | 0.266 | 0.250 | 0.000 | 0.276 | 64.000 | -0.512 |
| poker spade | card | fire | ace | shovel | 2.150 | 0.272 | 0.000 | 0.070 | 0.226 | 65.000 | -0.418 |

Supplemntary Information S2: Word Pairs with Priming

Results of the experiment by

Bruza, P.D., Kitto, K., Ramm, B.J., & Sitbon, L.,

analyzed in Section 6 of the main text;

all the computations are made by Bruza et al., except for the last column

and coversion of probabilities into expectations

Reproduced with permission.

# Supplementary Information S3:
# Details of the Matching Experiment

## 1   Participants

All the participants were students at Purdue University. The second author of this paper, labeled as P3, participated in all the experiments. Two persons (P1 and P2) participated in Experiments 1(a) and 2(a), and two other persons (P4 and P5) in Experiments 1(b), 2(b), 2(c), 3(a), and 3(b). All participants were aged around 25 and had normal or corrected to normal vision.

## 2   Stimuli and Procedure

Visual stimuli consisting of curves and (sometimes) dots were presented on a flat-panel monitor. They were grayish-white on a comfortably low intensity background. The diameter of the dots and the width of the curves was 5 pixels (px). The participants viewed the stimuli in darkness using a chin rest with a forehead support at the distance of 90 from the monitor, making 1 screen pixel approximately 62 sec arc. In each trial the participants were asked to match a fixed stimulus by adjusting a variable stimulus by rotating a trackball using their dominant hand. Once a response was made to the participant's satisfaction, she or he clicked a button on the trackball device to end this trial, and a new stimulus appeared half a second later. Each experiment took several days, each of which consisted of about 200 trials conducted with a 10-min break in the middle; each such session was preceded by a practice series of 10 trials (which were not recorded).

## 2.1 Experiment 1(a)

Each trial began with presenting two circles with a dot in the first quadrant of each circle (as shown in Figure 1, top panels). The radius of each circle was 160 px. The dot in the upper left circle was fixed at one of randomly chosen six positions. Using the center of its circle as the origin, they can be represented equivalently using the rectangular coordinates: {(24 px, 48 px), (32 px, 32 px), (32 px, 64 px), (48 px, 24 px), (64 px, 32 px), (64 px, 64 px)} or the polar coordinates: {(53.67 px, 63.43 deg), (45.25 px, 45 deg), (71.55 px, 63.43 deg), (53.67 px, 26.57 deg), (71.55 px, 26.56 deg), (90.51 px, 45 deg)}. Hence the experimental design contained a $2 \times 2$ "rectangular" subdesign, {32 px, 64 px}×{32 px, 64 px}, and a $2 \times 2$ "polar" subdesign {53.67 px, 71.55 px}×{63.43 deg, 26.57 deg}.

The position of the dot in the bottom right circle was controlled by the trackball, until its location matched that of the fixed one. Once a response was made, the program recorded the locations of the target dot and the matching dot in both rectangular coordinates and polar coordinates. There were 1200 trials overall with approximately 200 trials per treatment.

## 2.2 Experiment 1(b)

The horizontal coordinate and vertical coordinate of the target dot were random integers drawn before each trial from the the rectangle [20 px, 80 px)×[20 px, 80 px). This Cartesian rectangle contained the polar-coordinate rectangle [40 px, 90 px)×[30 deg, 60 deg), allowing us to analyze the data falling within it separately. The overall number of trials was 1800, of which 900 fell within the polar-coordinate rectangle. In all other respect, the procedure was the same as in Experiment 1(a).

## 2.3 Experiment 2(a)

Each trial began as shown in Figure 1, middle left panel. The target figure, on the left, consisted of two concentric circles together with their center. The radii of circle 1 and circle 2 were randomly chosen from the sets {16 px, 56 px, 64 px} and {48 px, 72 px, 80 px}, respectively, in a $3 \times 3$ factorial

design. On the right, in the beginning of the trial, there was a dot located at (250 px, 0 px) relative to the center of the target figure. By rotating the trackball the participant aimed at matching the target figure by "blowing up" two circles from the dot on the right, one by one. Once the first matching circle was produced (inner or outer, the person could choose), the participant clicked a button on the trackball to stabilize this circle and then the program enabled him or her to "blow" the other circle. After the second match was made, the trial was terminated by clicking the same button on the trackball. The program recorded the radii of the target and matching concentric circles in each trial. There were 1800 trials overall, approximately 200 trials per treatment.

## 2.4   Experiment 2(b)

Experiment 2(b) was identical to Experiment 2(a) except that in each trial the radii of the target circle 1 and circle 2 were randomly chosen from four possibilities $\{12 \text{ px}, 24 \text{ px}\} \times \{18 \text{ px}, 30 \text{ px}\}$. There were 1600 trials overall, about 400 trials per treatment.

## 2.5   Experiment 2(c)

Experiment 2(c) was identical to Experiment 2(a) except that in each trial the radius of the target circle 1 was a number randomly chosen from the uniform distribution on the interval [18 px, 48 px) and the radius of the target circle 2 was randomly chosen from the interval [56 px, 86 px). There were 1800 trials overall.

## 2.6   Experiment 3(a)

Examples of two floral shapes together with their centers are shown in Figure 1, bottom panels. Two such configurations were presented simultaneously in each trial. The target one was on the left, the variable one on the right. The floral shape was generated using the function

$$x = cos(.02\pi\Delta)[70 + \alpha cos(.06\pi\Delta) + \beta cos(.1\pi\Delta)], \qquad (1)$$

$$y = sin(.02\pi\Delta)[70 + \alpha cos(.06\pi\Delta) + \beta cos(.1\pi\Delta)],$$

where $x, y$ stand for the rectangular coordinates. Amplitude $\alpha$ and amplitude $\beta$ of the left floral shape were randomly chosen from the sets {-18 px, 10 px, 14 px} and {-16 px, -12 px, 20 px}, respectively. $\Delta$ was varied from 0 to 99 with an increment of 1 at each step. At each step, a point with coordinates $(x, y)$ was drawn to the screen and each floral shape was composed of 100 such points. The $\alpha$ and $\beta$ for the shape on the right was controlled by rotating the trackball. The program converted the horizontal (vertical) component of the rotation to the change of $\alpha$ (respectively, $\beta$). The initial values for these amplitudes were randomly picked from the interval [-35 px, 35 px). There were 1800 trials overall, about 200 trials per treatment.

## 2.7  Experiment 3(b)

Experiment 3(b) was identical to Experiment 3(a) except that the two amplitudes for the target shape were randomly chosen numbers from the interval [-30 px, 30 px).

# 3  Results

The data are available as "Contextuality in Psychophysical Matching" dataset (Excel files) in http://dx.doi.org/10.7910/DVN/OJZKKP, Harvard Dataverse, V1.

# 4  Analysis

In each experiment we deleted outliers, defined, rather informally, as the matching values that were too far from the target values. The outliers made less than 1% of all data. (Note that the files in the repository do not have the outliers deleted.)

In Experiment 2b the design was $2 \times 2$. In Experiment 1a there were two $2 \times 2$ subdesigns, the

"rectangular" and "polar" ones. In Experiment 2a and 3a the design was $3 \times 3$ and the analysis was made for each of the nine possible $2 \times 2$ subdesigns. In Experiments 1b, 2c, 3b the values of the target stimulus were first dichotomized into below-median and above-median values, forming a $2 \times 2$ factorial design in each of them.

Once a $2 \times 2$ design was formed, the responses (matching values of the variable stimuli) were dichotomized as described in Section 7 of the main text, by choosing all possible combinations of four integer values in the intervals

$$x_1 \in \{\max[\min X_{12}, \min X_{41}], \min[\max X_{12}, \max X_{41}]\},$$
$$x_3 \in \{\max[\min X_{23}, \min X_{34}], \min[\max X_{23}, \max X_{34}]\},$$
$$y_2 \in \{\max[\min Y_{12}, \min Y_{23}], \min[\max Y_{12}, \max Y_{23}]\},$$
$$y_4 \in \{\max[\min Y_{41}, \min Y_{34}], \min[\max Y_{41}, \max Y_{34}]\}$$

The analysis afterwards consisted in computing the value of $\Delta C$. The number of dichotomizations in each $2 \times 2$ (sub)design was between 3024 and 11,663,568.



\end{document}